\pdfoutput=1 
\documentclass[aps,pra, amsmath,superscriptaddress, longbibliography,twocolumn, showkeys, floatfix, 10pt]{revtex4-2}
\usepackage[pdftex]{graphicx}
\usepackage{bm}
\usepackage{setspace}
\usepackage{indentfirst}
\usepackage{hyperref}  % uncomment for arxiv 
\usepackage{makecell}
\usepackage{xcolor}
\usepackage{xr-hyper}
\begin{document} 
\title{Modifying the Optical Emission of Vanadyl Phthalocyanine via Molecular Self-Assembly on van der Waals Materials}

\author{S. Carin Gavin}
\thanks{These authors contributed equally}
\affiliation{Department of Physics and Astronomy, Northwestern University, Evanston, IL 60208, USA}

\author{William Koll}
\thanks{These authors contributed equally}
\affiliation{Department of Physics, The Ohio State University, Columbus, Ohio 43210, USA}

\author{Moumita Kar}
\thanks{These authors contributed equally}
\affiliation{Department of Chemistry, Northwestern University, Evanston, IL 60208, USA}

\author{Yiying Liu}
\affiliation{Department of Physics and Astronomy, Northwestern University, Evanston, IL 60208, USA}

\author{Anushka Dasgupta}
\affiliation{Department of Materials Science and Engineering, Northwestern University, Evanston, IL 60208, USA}

\author{Ethan Garvey}
\affiliation{Department of Physics and Astronomy, Northwestern University, Evanston, IL 60208, USA}

\author{Thomas W. Song}
\affiliation{Department of Materials Science and Engineering, Northwestern University, Evanston, IL 60208, USA}

\author{Chunxi Zhou}
\affiliation{Department of Physics and Astronomy, Northwestern University, Evanston, IL 60208, USA}

\author{Brendan P. Kerwin}
\affiliation{Department of Chemistry, Northwestern University, Evanston, IL 60208, USA}

\author{Jash Jain}
\affiliation{Department of Physics and Astronomy, Northwestern University, Evanston, IL 60208, USA}

\author{Tobin J. Marks}
\affiliation{Department of Chemistry, Northwestern University, Evanston, IL 60208, USA}
\affiliation{Department of Materials Science and Engineering, Northwestern University, Evanston, IL 60208, USA}
\affiliation{Department of Chemical and Biological Engineering, Northwestern University, Evanston, IL 60208, USA}
\affiliation{Materials Research Center, Northwestern University, Evanston, IL 60208, USA}

\author{Mark C. Hersam}
\affiliation{Department of Chemistry, Northwestern University, Evanston, IL 60208, USA}
\affiliation{Department of Materials Science and Engineering, Northwestern University, Evanston, IL 60208, USA}
\affiliation{Materials Research Center, Northwestern University, Evanston, IL 60208, USA}
\affiliation{Department of Electrical and Computer Engineering, Northwestern University, Evanston, IL 60208, USA}

\author{George C. Schatz} 
\affiliation{Department of Chemistry, Northwestern University, Evanston, IL 60208, USA}

\author{Jay A. Gupta}
\affiliation{Department of Physics, The Ohio State University, Columbus, Ohio 43210, USA}

\author{Nathaniel P. Stern}
\email[]{n-stern@northwestern.edu}
\affiliation{Department of Physics and Astronomy, Northwestern University, Evanston, IL 60208, USA}

%\date{\today}

\begin{abstract}

Vanadyl phthalocyanine (VOPc) is a promising organic molecule for applications in quantum information because of its thermal stability, efficient processing, and potential as a spin qubit. The deposition of VOPc in different molecular orientations allows the properties to be customized for integration into various devices. However, such customization has yet to be fully leveraged to alter its intrinsic properties, particularly optical emission. Normally, VOPc films on dielectric substrates emit a broad photoluminescence peak in the near-infrared range, attributed to transitions in the Pc ring from its $\pi$ orbital structure. In this work, we demonstrate that the dominant optical transition of VOPc can be shifted by $\sim$ 250 meV through the controlled deposition of thin films on van der Waals material substrates. The weak interactions with van der Waals materials allow the molecules to uniquely self-assemble, resulting in modified optical behavior modulated by molecular phase and thickness. This work connects the self-assembling properties of molecules with their altered electronic structures and the resulting optical emission.

\end{abstract}
\maketitle

\section{Introduction}

Metallic phthalocyanines (Pcs) are organic molecules well known for applications in catalysis, photovoltaics, biological imaging and sensing, and thin-film microelectronics~\cite{melville2015phthalocyanine, cranston2021metal, li2008molecular}. Among them, vanadyl phthalocyanine (VOPc) has attracted particular interest due to its potential as a molecular spin qubit~\cite{atzori2016room, bonizzoni2017coherent, malavolti2018tunable, bader2016tuning, cimatti2019vanadyl}. The vanadium center in VOPc carries a stable unpaired electron spin ($S = 1/2$), offering spin coherence times on the order of microseconds at room temperature, while the molecule itself exhibits excellent chemical and thermal stability~\cite{atzori2016room}. Typically, a molecular spin qubit is most viable in a single isolated molecule to protect the electron spin from decoherence \cite{burkard2023semiconductor,noh2023vanadyl}. However, VOPc molecules retain their $S = 1/2$ properties as thin films and are relatively robust against changes in spin properties due to thickness or substrate \cite{cimatti2019vanadyl,xu2022orienting,lee2024interpreting, noh2023template}. This robustness presents a key advantage for integration into solid-state qubit platforms, particularly optical readout of spin states \cite{sun2018cavity}. As such, understanding and controlling the optical properties of VOPc thin films are important steps towards scalable spin-photon interfaces and other applications that exploit these molecules.

\begin{figure*}[tbph]
    \centering
    \includegraphics[scale=0.18]{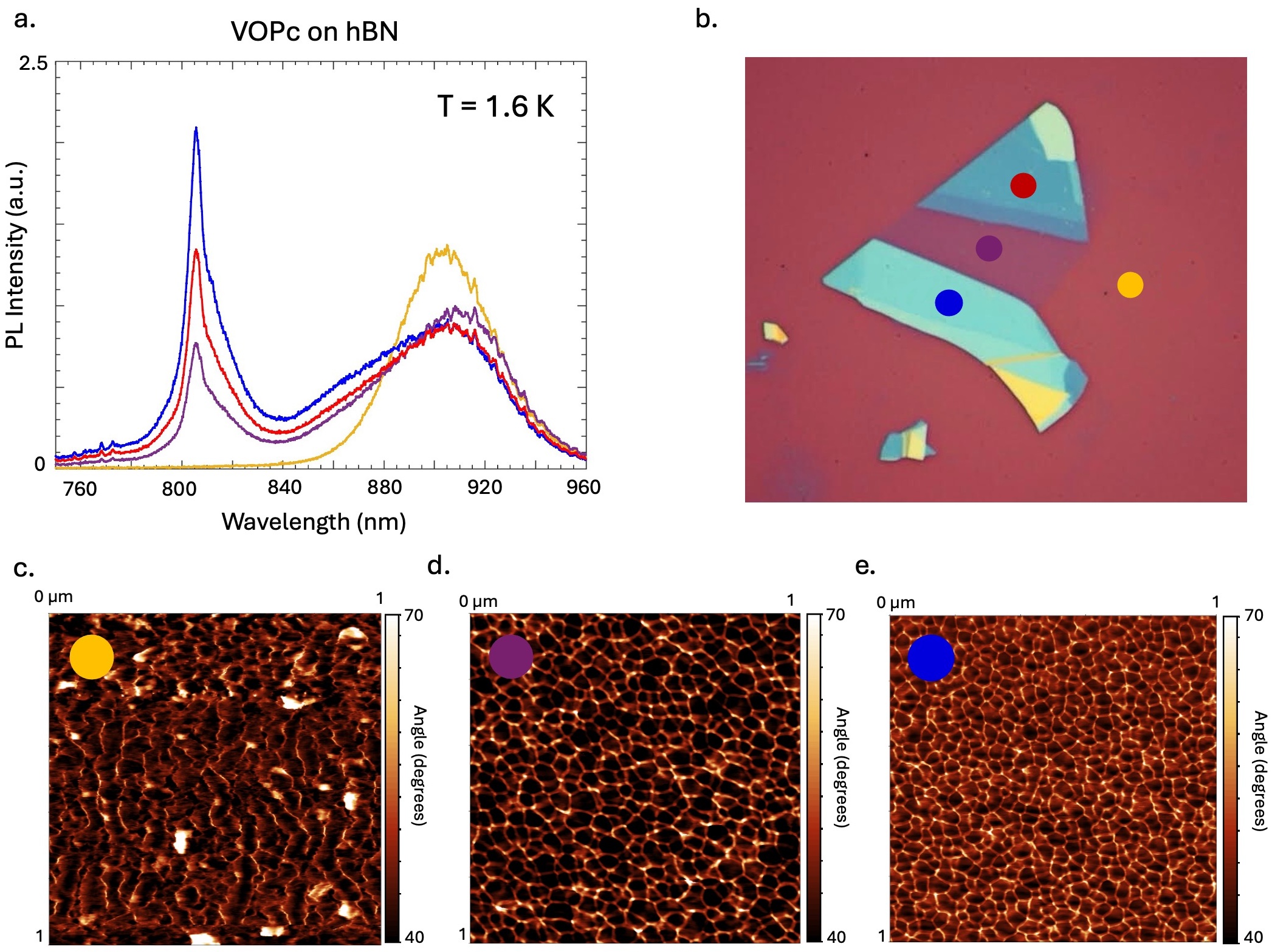}
    \caption{Novel emission from 5~nm VOPc on a multilayered flake of hBN. (a) PL spectra of the VOPc/hBN heterostructure in four different regions: bare Si/$\text{SiO}_\text{2}$ substrate (yellow), thickest hBN (blue), thinner hBN (red), and thinnest hBN (purple), corresponding to the regions pictured in the optical image (OM) of (b). On bare substrate, the VOPc emits only the characteristic broad PL around 900~nm. However, on hBN, a novel peak emerges around 805~nm, with an intensity trend that increases with hBN thickness. (c)-(f) Phase contrast atomic force microscopy (AFM) of VOPc on the regions corresponding to colors in Figure (b). On bare substrate, the molecular layer has less well-defined domains, whereas on thin (purple) and thick (blue) hBN, the domains become small and more defined, indicating an overall change in morphology between substrate and vdW material.}
    \label{vopc-1}
\end{figure*}

Although the spin properties of VOPc are robust, other characteristics are highly sensitive to film morphology; variations in crystallinity, molecular orientation, phase composition, and alignment can significantly influence its optical absorption, charge transport, and electronic structure~\cite{collins1993optical, klyamer2020vanadyl, wang2014polymorphism, fang1996nonlinear, papageorgiou2004physics, yu2007optimizing, pakhomov2010phthalocyanine, li2008molecular, eguchi2013molecular}. VOPc is in a unique subcategory of metallic Pcs known as metal oxide Pcs, resulting from the oxygen molecule bonded out-of-plane of the Pc ring ~\cite{amsterdam2019electronic,cranston2021metal}. The protrusion of the vanadyl group from the Pc macrocycle naturally introduces two possible orientations with respect to a substrate: one with oxygen pointing above the molecular plane (“O-up”) and one below (“O-down”)~\cite{blowey2018structure, eguchi2013molecular, niu2014molecular, malavolti2018tunable}. In addition, VOPc exhibits well-known polymorphism, with two main crystalline phases—$\alpha$ and $\beta$—corresponding to triclinic and monoclinic lattice structures, respectively~\cite{griffiths1976polymorphism, ziolo1980phaseii}. This inherent structural diversity plays a crucial role in the optoelectronic properties of VOPc. Previous studies have shown that modifying substrate interfaces can tune molecular orientation, allowing enhanced charge transport in organic field-effect transistors~\cite{li2008molecular} and modulation of the electronic structure of VOPc at interfaces~\cite{eguchi2013molecular}. Although variations in film morphology have been leveraged for improved electronic transport properties \cite{wang2014polymorphism,ramadan2016morphology}, they have yet to be exploited to control the optical emission of VOPc thin films. 

An emerging approach for controlling molecular film properties is to interface them with atomically thin van der Waals (vdW) materials \cite{amsterdam2020tailoring}; these layered materials are versatile, customizable by stacking, free of dangling bonds, and excel in optoelectronic integration \cite{geim2013van,blackstone2021van,novoselov20162d}. A combination of molecular films and vdW materials creates atomically thin, mixed-dimensional heterostructures with tailored properties based on the constituent layers that can be easily integrated into various device architectures \cite{jariwala2017mixed,padgaonkar2019molecular,padgaonkar2020emergent,li2020molecular,amsterdam2021leveraging,utama2023mixed}. Understanding the optical features of these heterostructures is important for considering them for optoelectronic and spintronic applications. In this work, we demonstrate that depositing VOPc thin films ($\sim$ 5~nm) on vdW materials produces a new photoluminescence (PL) feature at higher energy than native emission of VOPc on typical dielectric substrates. The spectral position of the novel peak is consistent across four different vdW materials: molybdenum dilsufide ($\text{MoS}_\text{2}$), tungsten disulfide ($\text{WS}_\text{2}$), chromium iodide ($\text{CrI}_\text{3}$), and hexagonal boron nitride (hBN), indicating that its origins are independent of the band gap and magnetic ordering of the material. Reducing the thickness of the VOPc film to a single ``bilayer'' defined by an O-up, O-down self-assembly further blueshifts the dominant optical emission when deposited onto vdW materials, resulting in PL that is approximately 8$\times$ narrower and $\sim$250 meV higher in energy than the emission of films on dielectric substrates. To understand the relationship between VOPc orientation and thickness in bilayer and thin film form, density functional theory (DFT) calculations were performed. This modeling suggests that the band gap of VOPc is modulated by bilayer thickness, which provides a possible interpretation of our observations. Combining spectroscopy, microscopy, and computation, this work offers insight into the effect of crystalline phase on the electronic structure and corresponding optical emission of VOPc/vdW heterostructures and reveals an accessible method for controlling optical properties. By interfacing molecules with vdW materials, novel film structures and properties are unlocked. 

\begin{figure*}[tbph]
    \centering
    \includegraphics[width=\textwidth]{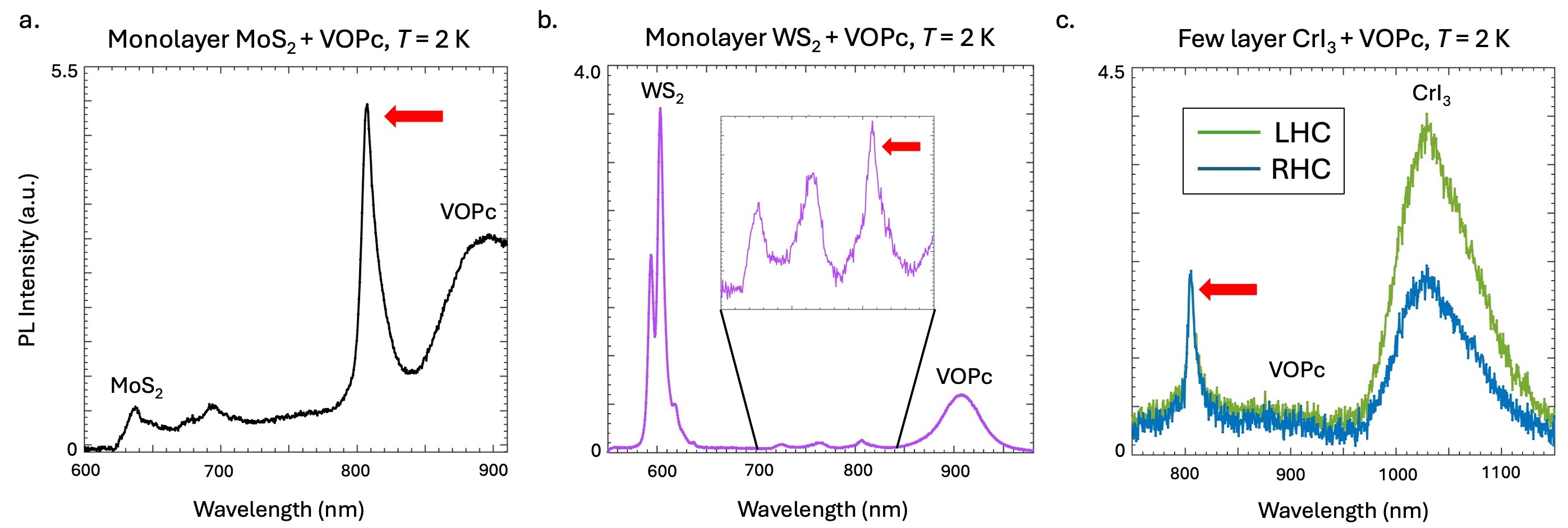}
    \caption{$T=2$~K spectra of (a) $\text{MoS}_\text{2}$, (b) $\text{WS}_\text{2}$, and (c) $\text{CrI}_\text{3}$ interfaced with VOPc thin films. Each spectrum has the native PL of the constituent components labeled, while the red arrow indicates the novel PL feature that appears only on the interface between materials. In (b), an inset is provided to more clearly see the interface spectral features, which are weak on the scale of $\text{WS}_\text{2}$ excitonic emission. In (c), the $\text{CrI}_\text{3}$/VOPc was assessed with circularly polarized PL; here the material is excited with left hand circularly (LHC) polarized light and measured for both right (blue) and left (green) hand polarization directions. The $\text{CrI}_\text{3}$ has a degree of circular polarization as expected, but VOPc and the interface emission do not.}
    \label{vopc-3}
\end{figure*}

\section{Optical Characterization of Thin Films}

Metal phthalocyanines are generally used in thin film form for scalability and optimized device performance \cite{cranston2021metal}. To assess the effect of substrate on VOP photoluminescence, thin films were deposited onto various vdW materials via thermal evaporation. The vdW materials were first mechanically exfoliated onto Si/$\text{SiO}_\text{2}$, which provided a direct comparison between the dielectric material and the vdW material for each sample. These films have a nominal thickness of 5~nm based on the quartz crystal mass reading of the evaporator. First, cryogenic optical spectroscopy of thin film VOPc on hexagonal boron nitride was performed. The laser excitation wavelength used for all hBN/VOPc results shown here is $\lambda$ = 656~nm; this wavelength is close to the peak absorption region of VOPc \cite{basova2008spectral,nanai1997polarized} and is not energetic enough to excite hBN \cite{wickramaratne2018monolayer}, meaning that all observed emission is directly related to the presence of VOPc. Figure \ref{vopc-1} shows the PL at $T=1.6$~K from a 5-nm thick VOPc film deposited on a multi-thickness flake of hBN. The yellow spectrum corresponds to VOPc that sits only on the Si/$\text{SiO}_\text{2}$ substrate, emitting the characteristic broad PL of a thin film \cite{schwinn2022charge,kong2022interlayer}. In contrast, for all regions of hBN, there is an additional emission feature centered around 805~nm. The intensity of this peak is correlated with hBN thickness, where the thinnest region (purple dot) exhibits the weakest PL, and the much thicker region (blue dot) shows the strongest. For all hBN regions, the native PL intensity is decreased in comparison to the bare substrate but is comparable across the hBN. In summary, a nominally 5~nm thick VOPc film on a dielectric substrate emits a single broad spectral feature in the near infrared (NIR) around 900~nm, but when deposited on hBN, a new optical transition emerges at higher energy that coexists with the original NIR emission. 

\begin{figure*}[tbph]
    \centering
    \includegraphics[width=0.75\textwidth]{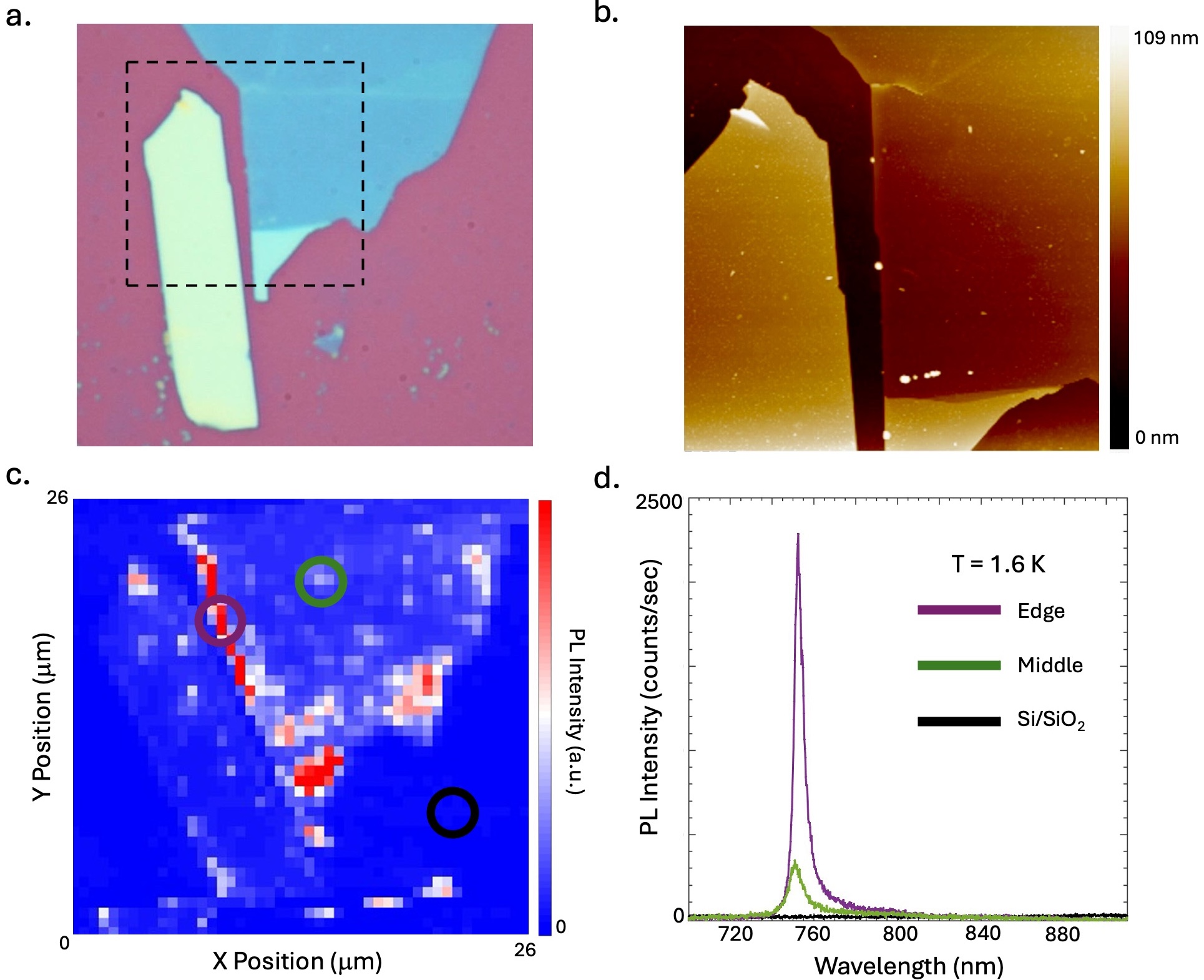}
    \caption{Optical properties of bilayer VOPc on hBN. (a) OM of the hBN flake with bilayer VOPc deposited on top. The black dashed box outlines the area scanned in the AFM of (b), which shows a relatively uniform height across the different thickness regions. (c) PL intensity map of the hBN area shown in the OM of (a). The PL is brighter near the edges of thinner hBN. Circles indicate the position of the spectra shown in (d), which shows a bright emission feature centered around 755~nm for film on hBN. Here, no PL is seen in the $\sim$900~nm region; rather, the new highest energy optical transition is preferred.}
    \label{vopc-2}
\end{figure*}

For insight into the origin of this novel PL, phase contrast atomic force microscopy (AFM) was used to image the sample structure, shown in Figures \ref{vopc-1}(c)-(e). Phase contrast AFM reveals information about the domain size and shape of the molecular film and helps differentiate ordering even when the overall height is uniform (Figure \ref{supp-heightAFM}). For films on a bare Si/$\text{SiO}_\text{2}$, the molecular domains have sharp, ragged edges (Figure \ref{vopc-1}c). In comparison, the film deposited on hBN shows smaller and more well-defined domains (Figures \ref{vopc-1} d,e). This indicates that the overall molecular ordering and packing is altered by the vdW substrate. Further insight on crystalline phase and ordering is discussed in Section \ref{sec:bilayer}.

Since AFM phase information suggests that the appearance of the new optical feature is correlated with substrate and film morphology, VOPc films of $\sim$3-5~nm thickness were deposited on other vdW materials to assess whether the novel PL is sensitive to different vdW substrates with varied crystal structures, band gaps, and magnetic ordering. Figure \ref{vopc-3} shows the PL spectra of the films on three different semiconductor vdW materials: monolayer $\text{MoS}_\text{2}$, monolayer $\text{WS}_\text{2}$, and few-layer $\text{CrI}_\text{3}$ (an Ising ferromagnet \cite{tomarchio2021low}). Interestingly, both the native VOPc PL and the novel peak at 805~nm are present for all materials, in addition to the characteristic emission of the respective vdW substrate. This means that the same optical transition is created regardless of the thickness, relative band gap (smaller or larger than VOPc), or magnetic ordering of the underlying semiconductor. This observation indicates that the changes in optical characteristics are indeed a result of a change in VOPc molecules themselves rather than of the electronic interactions of the heterojunction with the vdW material. All vdW materials show both native VOPc PL and higher energy emission, with two additional emission peaks accompanying the 805~nm feature on $\text{WS}_\text{2}$; both are higher energy, at 765~nm and 725~nm (discussed further in Section \ref{sec:modelling}). Additionally, circularly polarized PL was performed on the $\text{CrI}_\text{3}$, which is expected to have a degree of circular polarization \cite{seyler2018ligand}. As seen in Figure \ref{vopc-3}c, $\text{CrI}_\text{3}$ exhibits circular polarization, with the green and blue spectra corresponding to left and right handed detection. However, the interfacial VOPc feature at 805~nm is not circularly polarized, further indicating that the emission is distinct from the native $\text{CrI}_\text{3}$ PL.
Together, these spectroscopic data demonstrate that a novel, blue-shifted optical transition of VOPc emerges when films are deposited on vdW materials, and the resulting PL coexists with the original emission, with properties generally independent of the substrate's thickness, band gap, and magnetic ordering. Materials characterization with AFM shows that the morphology of the thin film is altered by the substrate and plays a role in the optical properties. To gain further insight on how molecular ordering can affect emission, we exploited self-assembly of VOPc molecules to test the thinnest possible film in which molecular alignment on the substrate would be most influential.

\begin{figure*}[tbph]
    \centering
    \includegraphics[scale=0.2]{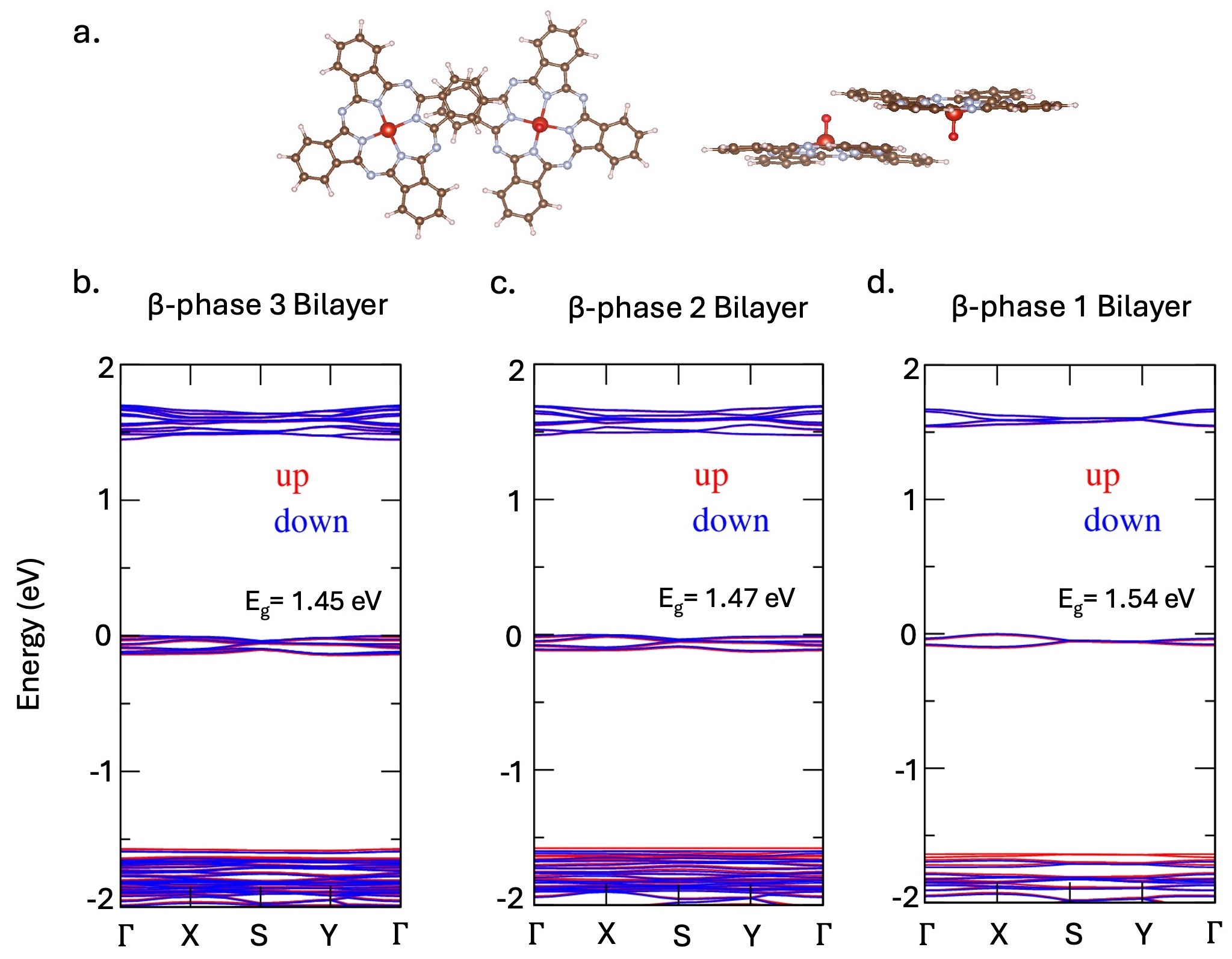}
    \caption{(a) The top-down and side view of beta phase molecules in bilayer formation. (b)-(d) Band structure of decreasing numbers of bilayers, showing an increasing band gap magnitude. Red represents spin up and blue represents spin down, and the valence band maximum has been shifted to zero. The magnitude difference between 3 and 2 layers is $\sim$20~meV, whereas the difference between 2 and 1 layers is $\sim$70~meV, indicating that the energy dependence is not linear but grows in a super-linear manner with decreasing layer number.}
    \label{modeling}
\end{figure*}

\section{VOPc Self-Assembly and Bilayer Optical Properties}\label{sec:bilayer}

Precise imaging of VOPc molecules on substrates is critical to directly connect orientation to optical properties. Koll \textit{et al.} used scanning tunneling microscopy (STM) to show that when a minimum thickness of VOPc is deposited onto silver, the molecules uniquely self-assemble and create a single ``bilayer'' of O-up, O-down molecules \cite{koll2024formation}.  Further studies by Kavand \textit{et al.} suggest that bilayer self-assembly extends beyond silver and is also the preferred formation when a bilayer film is deposited on hBN \cite{kavand2024general}. Following this literature precedent, bilayer VOPc films ($\sim$ 0.6~nm thick on average) were deposited onto multilayer hBN, and Figure \ref{vopc-2} shows the resulting cryogenic spectra. Figure \ref{vopc-2}a is a microscope image of the multilayer hBN, where the black dashed box outlines the portion assessed by height AFM in Figure \ref{vopc-2}b. The AFM indicates an overall smooth and uniform surface on the micron scale. Figure \ref{vopc-2}c is a PL intensity map of the hBN and substrate area covered by the bilayer film. The shape of the hBN flake is distinguishable in spatial imaging purely because of the VOPc emission, as the hBN itself does not emit with low power 656~nm laser excitation. Interestingly, the VOPc PL intensity is not uniform across the flakes. Rather, it preferentially emits near the edges of thinner hBN region. This observation may result from the tendency of Pc molecules to preferentially stabilize defect sites when adsorbed to a vdW material \cite{amsterdam2021mechanistic}, which is enabled by their diffusion across a substrate with which they have weak interactions \cite{papageorgiou2004physics,rana2022scanning, noh2023template}. This phenomenon is discussed in more detail in Section \ref{sec:discuss}. 

Figure \ref{vopc-2}d shows the PL spectra from the areas highlighted in the PL intensity map. For bilayer VOPc on an hBN substrate, native VOPc emission around 900~nm is not detected, and neither is the characteristic 805~nm features of the thicker films. Rather, emission is now observed around 750~nm. This PL peak is $\sim$8$\times$ narrower and roughly 260 meV higher in energy than that from VOPc emission on Si/$\text{SiO}_\text{2}$. The peak positions and widths are summarized in Figure \ref{supp-fits}. Notably, the bilayer emission feature is also detected at the same spectral position at room temperature, as seen in Figure \ref{supp-RTPL}a, although it is an order of magnitude weaker than at cryogenic temperatures. This observation differs from most molecular and crystalline semiconductor emission, which shifts with temperature due to vibrational interactions within the molecule or lattice \cite{furuyama2020temperature,kopaczek2022temperature}. For thicker VOPc films, the 805~nm peak at low temperatures is at the same spectral position as that of previous findings that report novel PL of VOPc films on $\text{MoS}_\text{2}$ at room temperature \cite{schwinn2022charge,kong2022interlayer}. Although VOPc emission on Si/$\text{SiO}_\text{2}$ shifts with temperature (Figure \ref{supp-RTPL}b), the modified features on vdW materials do not (Figure \ref{supp-RTPL}a) \cite{schwinn2022charge,kong2022interlayer}. This strongly suggests that molecular properties beyond the band gap have been altered by bilayer formation. While vibrational modes would normally be modified by changing temperature \cite{pan1998structure}, these new transitions do not exhibit typical spectral shifts.

\section{Modeling of Molecular Phases}\label{sec:modelling}

Given the abundance of experimental data indicating that VOPc packing structure and assembly affect its optical emission, first-principles calculations were performed to understand how those properties alter the electronic structure of the molecules. VOPc molecules have a characteristic band gap defined by their highest occupied molecular orbital (HOMO) and lowest unoccupied molecular orbital (LUMO) \cite{singh2010optical,baba2016electronic}. The calculated values from molecular simulations can often conflict with the experimental values since the molecules are strongly affected by substrate, morphology, and deposition method, as discussed in this work. A crucial determinant of VOPc thin film behavior is the crystalline phase of the molecular film, which includes both orientation and the relative overlap between neighboring molecules \cite{meletov2020phonon,xiong2024exciton}. 

First, DFT calculations of single molecules on hBN were performed to check whether the electronic structure is altered by the orientation of the oxygen molecule, thereby isolating whether O-up or O-down configurations alone determine the changes in optical emission. The results shown in Figure \ref{supp-Orientation} indicate that VOPc HOMO/LUMO levels are not altered by oxygen orientation of VOPc on hBN and would not alone account for the observed optical changes. However, calculating the energetic stability of each orientation reveals that O-up is the preferable orientation of molecules in direct contact with the hBN, with an energetic difference of about 0.4 eV. This is consistent with previous studies of sub-monolayer VOPc molecule orientation and the overall bilayer self-assembly model \cite{niu2014molecular,cimatti2019vanadyl,koll2024formation, padgaonkar2019molecular} which conclude that molecules closest to the substrate are O-up.

Because the relative orientation of oxygen to hBN itself does not alter the electronic structure, further calculations were performed to probe the effect of crystalline phase. Previous work has demonstrated that metallic Pcs adopt a face-on configuration (where the Pc plane lies flat on the vdW surface) when in contact with a vdW surface \cite{padgaonkar2019molecular}, and further experiments show that the face-on configuration can affect exciton dynamics. A recent study by Xiong \textit{et al.} examined the exciton dynamics of titanyl phthalocyanine deposited on sapphire substrates and the vdW semiconductor tungsten diselenide \cite{xiong2024exciton}. The spectroscopic results show that molecules in contact with tungsten diselenide adopt the face-on beta phase configuration and exhibit blue-shifted exciton emission. In our work, STM confirms that molecules are face-on with the vdW substrates \cite{koll2024formation}. Furthermore, Koll \textit{ et al.} found that the O-up/O-down bilayer configuration is more energetically stable when there is a large lateral offset between the oxygen of the sandwiched atoms \cite{koll2024formation}. In the beta phase, the lateral distance between oxygen atoms is 10.34 \AA, which is larger than the alpha phase distance of 7.46 \AA~(Figure \ref{supp-phases}). Because the beta phase is defined by the face-on configuration of the molecule with the substrate and has the larger oxygen offset, it is a likely candidate for our emission. However, our results show that the optical emission depends on the thickness of the film. Where the emission of a 5~nm film is altered by a vdW material, the dominant PL is further blue-shifted by reducing the film thickness to a single bilayer. If the beta-phase configuration is responsible, its electronic structure should be altered by layer number.

Indeed, the results of DFT show that the beta-phase band structure is altered by layer number, with the band gap widening as layer number is reduced. Figure \ref{modeling}a shows the schematic of the overlap and lateral offset of two VOPc molecules in the beta phase, and Figures \ref{modeling}(b-d) demonstrate the increase in the band gap with a decrease in the number of beta phase bilayers. When considered in 3D form (which can be considered infinite in all directions), the band gap is 1.42 eV, corresponding to emission at about 870~nm (Figure \ref{supp-3dDFT}). Typically, DFT underestimates solid-state band gaps due to exchange interactions and other factors \cite{wan2021effectively}, and we note that the calculated single bilayer magnitude of 1.54 eV (corresponding to 805~nm) does not precisely match the measured bilayer spectral position of 750~nm. Yet when the beta phase is modeled in its 2D form as sheets of bilayers, the band gap increases compared to the 3D form and with decreasing bilayer number. The experimental evidence and precedent from other investigations of metallic Pcs strongly suggest that the beta-phase orientation of molecules in the thin film is responsible for the varied spectral features.

\begin{figure*}[tbhp]
    \centering
    \includegraphics[width=\textwidth]{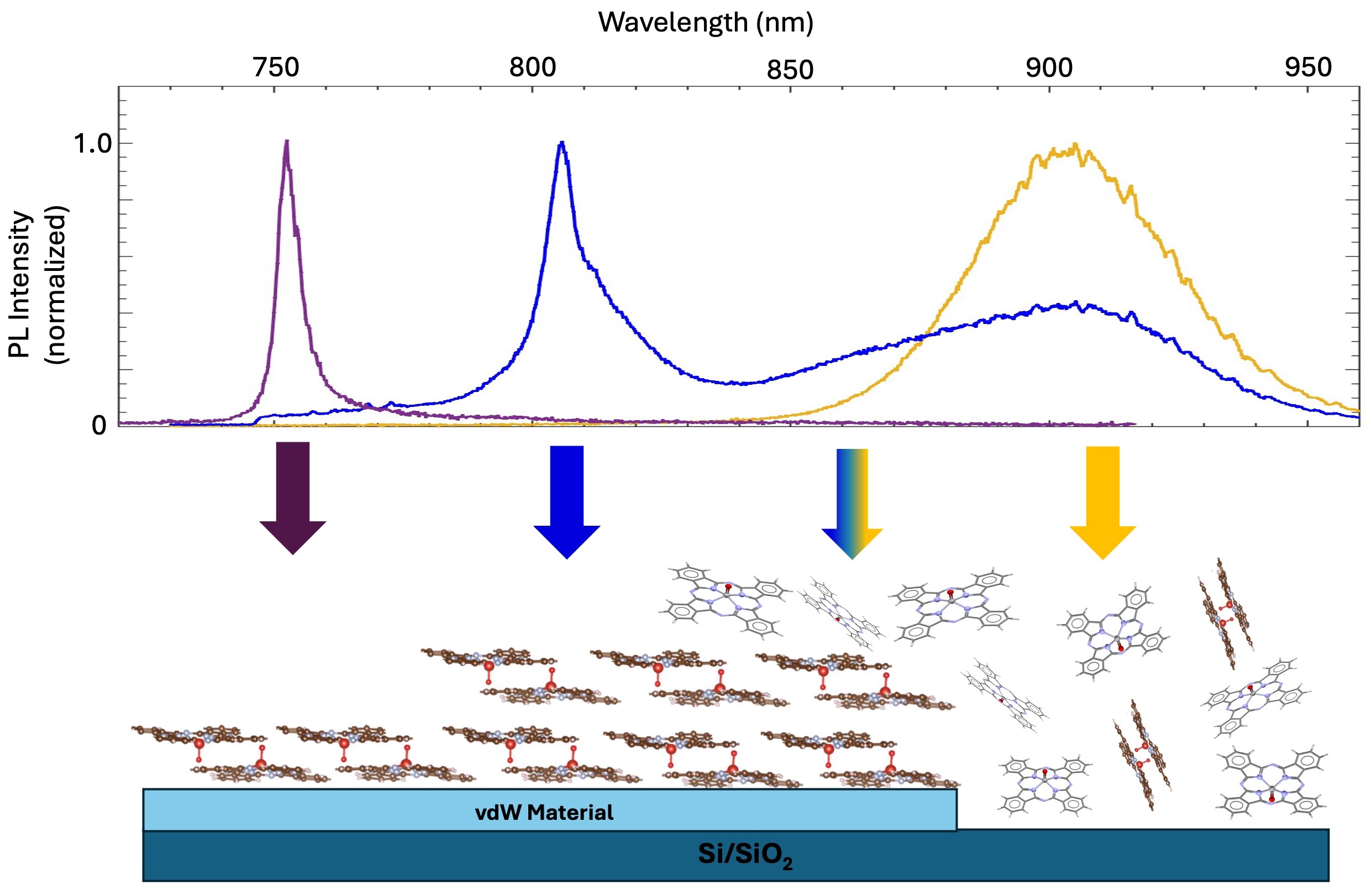}
    \caption{Summary of the optical emission based on ordering and layer number; all spectra have been normalized such that their dominant spectral feature is equal to one. The highest energy emission peak (purple) is the result of a single O-up/O-down beta-phase bilayer, enabled by weak interactions with hBN. As film thickness increases on hBN, ordered bilayer number increases (blue), and the novel PL shifts to lower energy from band gap reduction. The schematic for the dominant peak in the blue spectrum represents $>$1 bilayer generically rather than two bilayers specifically. Additionally, unordered molecules from top layers contributed to broad NIR PL. Finally, when a thicker film is deposited on only Si/$\text{SiO}_\text{2}$ substrate (yellow), the molecules are primarily unordered but may include edge-to-edge alpha phase molecules, resulting in only the characteristic NIR emission.}
    \label{summary}
\end{figure*}

Finally, time-dependent DFT (TDDFT) was used to perform excited-state calculations of both alpha- and beta-phase molecules, which reveals the oscillator strength of different transitions. Native emission from VOPc molecules is typically attributed to $\pi$-$\pi$ orbital transitions from the Pc ring \cite{zhang2007electronic, debnath2022triggering, adler2015interface}, with negligible contribution from the metal center. From the data presented in Table \ref{supp-table-tddft}, we observe that the excited states with high oscillator strengths also originate mainly from Pc ring-centered transitions (Pc-to-Pc transitions) for both formations. In contrast, metal-centered transitions exhibit very low oscillator strengths, and as such their contribution is negligible. Based on this, we conclude that the experimentally observed emission arises primarily from Pc-centered transitions. However, the calculated exciton energies are slightly overestimated, which could be due to the level of theory used, seen similarly in other works calculating such values for metallic Pcs \cite{theisen2015ground}. Furthermore, our model includes only two VOPc molecules (Figure \ref{supp-phases}), meaning that the excitation energies between the phases do not differ significantly. This limited variation is due to the small system size, and we expect that increasing the number of molecules would reveal a more pronounced difference between the two phases. Nonetheless, the TDDFT provides valuable insight into the atomic origin of the emission observed here. It has not been fundamentally altered; rather, the band gap and primary optical transition are tailored by the interaction of the Pc rings in beta-phase formation and with the number of bilayers in the system. This is critical for considering how novel emission is being modified to higher, rather than lower, energies. Transitions within the Pc rings are due to transitions between the HOMO/LUMO levels and are characterized by an absorption spectrum with a range of peaks, called the Q band \cite{wang2022photochemical,cao2022steady}. The Q band in VOPc and other metal Pcs has multiple absorption peaks \cite{van1996emission,edwards1970porphyrins}, but usually only has one characteristic emission feature that is the highest energy transition possible in the system. In order for new optical transitions to be at higher energies than this, the HOMO/LUMO band gap would need to increase. Furthermore, it provides a possible interpretation of observing not just one, but multiple higher energy emission features, as in the VOPc heterostructure with $\text{WS}_\text{2}$ (Figure \ref{vopc-3}b). Since the Q band encompasses multiple absorption peaks, orientation and phase may shift these peaks and activate more than one as an emissive transition. We note that in other metallic Pcs, the Q band absorption peaks are spaced 40~nm apart \cite{van1996emission}, and the emission peak spacing in $\text{WS}_\text{2}$ is also 40~nm with features at 725~nm, 765~nm, and 805~nm.

\section{Discussion}\label{sec:discuss}

In 2004, Papageorgiou \textit{et al.} extensively studied the adsorption geometry of phthalocyanines on various substrates, revealing through multiple techniques that metallic Pcs do not create ordered films on silicon-based substrates. However, for submonolayer through multilayer thicknesses, ordering emerges when deposited on vdW semiconductors \cite{papageorgiou2004physics}. Although the exact formation of the metallic Pc's strongly depends on the specific transitional metal and the deposition method, the takeaway is clear: weak molecule/vdW interactions enable molecular self-assembly where stronger interactions with dielectric silicon-based substrates do not. This picture has since been supported by many investigations \cite{eguchi2013molecular,veiga2016adsorption, jarvinen2013molecular,jarvinen2014self, amsterdam2019electronic, xiong2024exciton,noh2023template}. Here, we have revealed the unique properties of VOPc molecules based on this precedent. Their O-up/O-down bilayer configuration was confirmed via STM by Koll \textit{et al.}, and prior calculations suggested that the lateral offset plays a key role in the energetic stability of the bilayer. Our computational results support this picture; we demonstrate that indeed, O-up VOPc molecules are energetically preferred as the orientation immediately in contact with a vdW material. Furthermore, DFT reveals that the specific lateral offset between the O-up and O-down molecules in bilayer form not only matters for stability but also helps explain changes in the electronic structure. We show that in the beta phase, bilayer number modulates the band gap of VOPc, widening the gap as the number is reduced.

\begin{center}
\begin{table}[phb]
\begin{tabular}{c| c | c | c }
	\hline\hline
  \textbf{Phase} & \textbf{Thickness} & \textbf{Band gap (eV)} & \textbf{Emission (eV)}\\ [0.5ex]
  \hline
 Alpha & 3D & 1.15 & 1.37 \\ 
 \hline
 Beta & 3D & 1.42 & 1.54 \\
 \hline
 Beta & B-3L & 1.45 & - \\
 \hline
 Beta & B-2L & 1.47 & -  \\
 \hline
 Beta & B-1L & 1.54 & 1.64\\

 \end{tabular}
\vspace{-0em}
\caption{Summary of computational results of VOPc crystalline phase, modeled thickness, and band gap compared to the observed emission energies. Here, 3D refers to the films modeled as infinite in all directions, where B-\#L refers to a specific number of modeled bilayers. The smallest band gap comes from the alpha phase molecules, whereas the largest band gap is a single bilayer of beta phase molecules. The experimental emission energies are shown for 3D alpha phase (few-nm films on Si/$\text{SiO}_\text{2}$), 3D beta phase (few-nm films on vdW materials), and single bilayer beta phase (single bilayer on vdW materials). The change in emission energy between one and three bilayers is not differentiated experimentally in our observations.\vspace{0em}}\vspace{-0em}
\label{calc-summ}
\end{table}
\end{center}

In the previously cited work by Xiong \textit{et al.}, the authors show that alpha phase molecules in an edge-to-edge configuration (where the Pc plane creates a large angle with the sapphire substrate) are responsible for broad emission at the lowest energy \cite{xiong2024exciton}. However, they only observe this substrate feature after the sample is annealed, which encourages the molecules to self-assemble. On sapphire without annealing, the molecules are considered amorphous and do not emit. In comparison, the VOPc films on Si/$\text{SiO}_\text{2}$ presented here emit without annealing. However, as discussed in this section, it is unlikely that molecules in a relatively thick film on Si/$\text{SiO}_\text{2}$ have self-assembled. We therefore postulate that the broadest emission of VOPc films on dielectric substrates could be attributed to a combination of amorphous (unordered) and alpha phase molecules. A summary of the computational findings is shown in Table \ref{calc-summ}. VOPc molecules in the 3D alpha phase have the smallest band gap of all modeled systems, indicating their contribution to low energy emission. In the 3D beta phase, this band gap increases significantly, demonstrating that crystalline phase can be responsible for shifting optical emission to higher energies (shorter wavelengths), as observed in experiment. Furthermore, that band gap can be further widened by specifically reducing the number of beta phase bilayers in the system. The computed spectral shift between 3D beta phase molecules and a single bilayer is $\sim$120 meV, on the same order of magnitude as the experimental shift observed between the thicker film emission (805~nm) and single bilayer VOPc emission (750nm) on vdW materials, which is $\sim$102 meV (Table \ref{calc-summ}). The experimental relationship between substrate, thickness, and emission is represented in Figure \ref{summary}. Thicker films of VOPc molecules that do not self-assemble on Si/$\text{SiO}_\text{2}$ emit broadly in the near-infrared range around 900~nm. When several layers of molecules self-assemble in the beta phase near a vdW substrate, facilitated by weak interactions with the vdW material, a new PL peak emerges that exists alongside the original emission. Thicker layers of the vdW material increase the intensity of the novel peak as interactions between the molecules and dielectric substrate are reduced, facilitating further self-assembly (Figure \ref{vopc-1}a). When a minimum amount of VOPc molecules are deposited such that a single bilayer can form on a vdW material, the original NIR emission and the interfacial 805~nm feature are eliminated, with the new dominant optical transition at 750~nm. This feature is much narrower than VOPc emission in Si/$\text{SiO}_\text{2}$ and does not shift with temperature, both of which indicate that the influence of vibrational modes has been greatly reduced. With these experiments and calculations, we have illuminated the complexity of molecular photoluminescence and the importance of considering phase and layer interactions in its behavior. In doing so, a new control knob has been created for this versatile molecule.

\section{Methods}

\subsection{Low Temperature Optical Spectroscopy}

Cryogenic optical spectroscopy was performed in an AttoDry 2100 cryostat at $T=1.6$~K and a Quantum Design Opticool cryostat at $T=2$~K. High resolution spatial mapping of VOPc/hBN was performed using Attocube piezoeletric nanopositioners in the $x-y$ plane within the AttoDry cryostat. Confocal spectroscopy of VOPc/TMDs was performed using a 0.82 NA, $100\times$ magnification objective with a 532 or 656 nm diode laser excitation. The diffraction-limited spot sizes are estimated to be $D=1.22\lambda/\text{NA} \approx 0.79$~$\mu$m and $D=1.22\lambda/\text{NA} \approx 0.98$~$\mu$m, respectively. Light was collected with an optical fiber and sent to a 750-mm focal length spectrometer (Andor Shamrock SR-750) with a thermoelectrically cooled CCD camera (DU420A-BEX2-DD). Confocal spectroscopy of VOPc on $\text{CrI}_\text{3}$ was performed with a 633 nm continuous wave laser excitation focused using a Nikon 100$\times$ 0.6 NA long working distance objective. The diffraction-limited spot size is estimated to be $D=1.22\lambda/\text{NA} \approx 1.29$~$\mu$m. Photoluminescence was collected by the same objective and measured with a Shamrock 303i spectrometer with a thermoelectrically cooled CCD camera (DU420A-BEX2-DD).

\subsection{Sample Design and Molecular Deposition}

For creating bilayer films, an hBN flake was first acquired via micromechanical exfoliation and subsequent transfer onto an Si/$\text{SiO}_\text{2}$ substrate, which was then introduced to an ultra-high vacuum (UHV) chamber (base pressure $~10^{-10}$  mbar) and de-gassed at 100°C for 1 hour. VOPc powder was synthesized by the Freedman lab at MIT via an adapted literature procedure\cite{tomoda_synthesis_1983}. The powder was loaded into an alumina-coated heater basket, and de-gassed at 150°C for 12 hours under high vacuum conditions ($10^{-8}$ mbar). One bilayer of VOPc was sublimated at 200°C under UHV conditions onto the substrate maintained at room-temperature. The deposition rate ($\sim$1 monolayer/min) was previously calibrated by STM imaging of VOPc deposited on Ag(100)\cite{koll2024formation}. 

For thicker films, VOPc powder of purity ($>90\%$) was purchased from Sigma-Aldrich. Purification of the VOPc powder was performed via thermal gradient sublimation. The commercial VOPc powder was heated in a tube furnace at 400°C and 70 mTorr. A temperature gradient of 100°C was maintained to allow the VOPc to recrystallize further downstream.  A 5-nm VOPc film was then grown on the exfoliated 2D flakes using the sublimated powder with a thermal evaporator housed inside an inert N2 glove box at a rate of 0.1 \AA/sec. 

\subsection{Modeling}

All calculations were performed using the Vienna Ab initio Simulation Package (VASP), DFT, employing a plane-wave basis set and the projector-augmented wave (PAW) method \cite{kresse1999ultrasoft}. Geometry optimizations were performed using the generalized gradient approximation (GGA) with the Perdew–Burke–Ernzerhof (PBE) functional \cite{paier2005perdew}. Grimme’s DFT-D3 dispersion correction was employed to account for van der Waals interactions \cite{grimme2010consistent}. The energy convergence criterion was set to 0.0001 eV, and for geometry optimization, the atoms were relaxed until the force on each atom was less than 0.001 eV/\AA. An energy cutoff of 450 eV was used for the plane-wave basis set. The Brillouin zone was sampled using a $4\times4\times1$ Monkhorst–Pack k-point grid for two-dimensional systems and a $4\times4\times4$ grid for three-dimensional systems. To avoid interactions between periodic images in the two-dimensional systems, a vacuum spacing of approximately 15 \AA\, was applied along the c-axis. All band structure calculations along the high-symmetry paths of the Brillouin zone were performed using the range-separated Heyd–Scuseria–Ernzerhof hybrid functional (HSE06), based on geometries optimized with the PBE-D3 functional \cite{heyd2003hybrid}.  Band structures and other post-processing analyses were performed using VASPKIT \cite{wang2021vaspkit}.

To investigate the nature of the excited-state properties, we performed TDDFT calculations using the Turbomole 7.7 package \cite{di2016u}. The TDDFT calculations were performed using the B3LYP functional and the def-SVP basis set. We also tested other functionals, such as CAM-B3LYP and $\omega$B97XD, as well as a larger basis set like def-TZVP, using a single VOPc molecule. However, the B3LYP functional with the def-SVP basis set provided reasonably accurate results, which is why we adopted this combination for the larger system. To account for van der Waals interactions, the DFT-D3 dispersion correction proposed by Grimme was applied. To study the excited-state properties of alpha and beta phases, we considered a molecular model containing two VOPc molecules.

\section*{Author Contributions}

S.C.G., Y.L., and E.G. performed cryogenic optical spectroscopy. W.K. deposited bilayer VOPc films; A.D. and T.W.S. deposited 5~nm VOPc films that were sublimated by B.P.K. M.K. executed first-principles calculations. J.J. analyzed spectral data. A.D. and C.Z. performed materials characterization. B.K. acknowledges support from the U.S. Department of Commerce, National Institute of Standards and Technology (Award 70NANB19H005) as part of the Center for Hierarchical Materials Design (CHiMaD). S.C.G. prepared the manuscript with the assistance of Y.L. and contributions from all authors. T.J.M., M.C.H., G.C.S., J.A.G., and N.P.S. supervised the project.

\section*{Competing Interests}

The authors declare no competing interests.

\begin{acknowledgments}

This research was primarily supported as part of the Center for Molecular Quantum Transduction, an Energy Frontier Research Center funded by the U.S. Department of Energy (DOE), Office of Science, Basic Energy Sciences (BES), under Award No. DE-SC0021314. W.K. and J.A.G. recognize support from NSF NQVL:QSTD:PILOT MPS-
2435363. Aspects of this work (assembly of the mixed dimensional Pc/vdW heterostructures) were partially supported by the National Science Foundation’s MRSEC program (DMR-
2308691) at the Materials Research Center of Northwestern University. This work made use of the MLTOF facility which has received support from the MRSEC Program (NSF DMR-2308691) of the
Materials Research Center at Northwestern University and from Northwestern University. S.C.G. extends her thanks to Prof. Ezekiel Johnston-Halperin, who provided many helpful discussions about molecules, and Dr. Rianna Greer in the lab of Prof. Danna Freedman at MIT for synthesizing VOPc powders used by W.K. and J.A.G.

\end{acknowledgments}

\clearpage

\bibliography{citations}

\begin{thebibliography}{75}%
\makeatletter
\providecommand \@ifxundefined [1]{%
 \@ifx{#1\undefined}
}%
\providecommand \@ifnum [1]{%
 \ifnum #1\expandafter \@firstoftwo
 \else \expandafter \@secondoftwo
 \fi
}%
\providecommand \@ifx [1]{%
 \ifx #1\expandafter \@firstoftwo
 \else \expandafter \@secondoftwo
 \fi
}%
\providecommand \natexlab [1]{#1}%
\providecommand \enquote  [1]{``#1''}%
\providecommand \bibnamefont  [1]{#1}%
\providecommand \bibfnamefont [1]{#1}%
\providecommand \citenamefont [1]{#1}%
\providecommand \href@noop [0]{\@secondoftwo}%
\providecommand \href [0]{\begingroup \@sanitize@url \@href}%
\providecommand \@href[1]{\@@startlink{#1}\@@href}%
\providecommand \@@href[1]{\endgroup#1\@@endlink}%
\providecommand \@sanitize@url [0]{\catcode `\\12\catcode `\$12\catcode `\&12\catcode `\#12\catcode `\^12\catcode `\_12\catcode `\%12\relax}%
\providecommand \@@startlink[1]{}%
\providecommand \@@endlink[0]{}%
\providecommand \url  [0]{\begingroup\@sanitize@url \@url }%
\providecommand \@url [1]{\endgroup\@href {#1}{\urlprefix }}%
\providecommand \urlprefix  [0]{URL }%
\providecommand \Eprint [0]{\href }%
\providecommand \doibase [0]{https://doi.org/}%
\providecommand \selectlanguage [0]{\@gobble}%
\providecommand \bibinfo  [0]{\@secondoftwo}%
\providecommand \bibfield  [0]{\@secondoftwo}%
\providecommand \translation [1]{[#1]}%
\providecommand \BibitemOpen [0]{}%
\providecommand \bibitemStop [0]{}%
\providecommand \bibitemNoStop [0]{.\EOS\space}%
\providecommand \EOS [0]{\spacefactor3000\relax}%
\providecommand \BibitemShut  [1]{\csname bibitem#1\endcsname}%
\let\auto@bib@innerbib\@empty
%</preamble>
\bibitem [{\citenamefont {Melville}\ \emph {et~al.}(2015)\citenamefont {Melville}, \citenamefont {Lessard},\ and\ \citenamefont {Bender}}]{melville2015phthalocyanine}%
  \BibitemOpen
  \bibfield  {author} {\bibinfo {author} {\bibfnamefont {O.~A.}\ \bibnamefont {Melville}}, \bibinfo {author} {\bibfnamefont {B.~H.}\ \bibnamefont {Lessard}},\ and\ \bibinfo {author} {\bibfnamefont {T.~P.}\ \bibnamefont {Bender}},\ }\bibfield  {title} {\bibinfo {title} {Phthalocyanine-based organic thin-film transistors: a review of recent advances},\ }\href@noop {} {\bibfield  {journal} {\bibinfo  {journal} {{ACS} {A}pplied {M}aterials \& {I}nterfaces}\ }\textbf {\bibinfo {volume} {7}},\ \bibinfo {pages} {13105} (\bibinfo {year} {2015})}\BibitemShut {NoStop}%
\bibitem [{\citenamefont {Cranston}\ and\ \citenamefont {Lessard}(2021)}]{cranston2021metal}%
  \BibitemOpen
  \bibfield  {author} {\bibinfo {author} {\bibfnamefont {R.~R.}\ \bibnamefont {Cranston}}\ and\ \bibinfo {author} {\bibfnamefont {B.~H.}\ \bibnamefont {Lessard}},\ }\bibfield  {title} {\bibinfo {title} {Metal phthalocyanines: {T}hin-film formation, microstructure, and physical properties},\ }\href@noop {} {\bibfield  {journal} {\bibinfo  {journal} {RSC {A}dvances}\ }\textbf {\bibinfo {volume} {11}},\ \bibinfo {pages} {21716} (\bibinfo {year} {2021})}\BibitemShut {NoStop}%
\bibitem [{\citenamefont {Li}\ \emph {et~al.}(2008)\citenamefont {Li}, \citenamefont {Tang}, \citenamefont {Li},\ and\ \citenamefont {Hu}}]{li2008molecular}%
  \BibitemOpen
  \bibfield  {author} {\bibinfo {author} {\bibfnamefont {L.}~\bibnamefont {Li}}, \bibinfo {author} {\bibfnamefont {Q.}~\bibnamefont {Tang}}, \bibinfo {author} {\bibfnamefont {H.}~\bibnamefont {Li}},\ and\ \bibinfo {author} {\bibfnamefont {W.}~\bibnamefont {Hu}},\ }\bibfield  {title} {\bibinfo {title} {Molecular orientation and interface compatibility for high performance organic thin film transistor based on vanadyl phthalocyanine},\ }\href@noop {} {\bibfield  {journal} {\bibinfo  {journal} {The {J}ournal of {P}hysical {C}hemistry {B}}\ }\textbf {\bibinfo {volume} {112}},\ \bibinfo {pages} {10405} (\bibinfo {year} {2008})}\BibitemShut {NoStop}%
\bibitem [{\citenamefont {Atzori}\ \emph {et~al.}(2016)\citenamefont {Atzori}, \citenamefont {Tesi}, \citenamefont {Morra}, \citenamefont {Chiesa}, \citenamefont {Sorace},\ and\ \citenamefont {Sessoli}}]{atzori2016room}%
  \BibitemOpen
  \bibfield  {author} {\bibinfo {author} {\bibfnamefont {M.}~\bibnamefont {Atzori}}, \bibinfo {author} {\bibfnamefont {L.}~\bibnamefont {Tesi}}, \bibinfo {author} {\bibfnamefont {E.}~\bibnamefont {Morra}}, \bibinfo {author} {\bibfnamefont {M.}~\bibnamefont {Chiesa}}, \bibinfo {author} {\bibfnamefont {L.}~\bibnamefont {Sorace}},\ and\ \bibinfo {author} {\bibfnamefont {R.}~\bibnamefont {Sessoli}},\ }\bibfield  {title} {\bibinfo {title} {Room-temperature quantum coherence and rabi oscillations in vanadyl phthalocyanine: toward multifunctional molecular spin qubits},\ }\href@noop {} {\bibfield  {journal} {\bibinfo  {journal} {Journal of the {A}merican {C}hemical {S}ociety}\ }\textbf {\bibinfo {volume} {138}},\ \bibinfo {pages} {2154} (\bibinfo {year} {2016})}\BibitemShut {NoStop}%
\bibitem [{\citenamefont {Bonizzoni}\ \emph {et~al.}(2017)\citenamefont {Bonizzoni}, \citenamefont {Ghirri}, \citenamefont {Atzori}, \citenamefont {Sorace}, \citenamefont {Sessoli},\ and\ \citenamefont {Affronte}}]{bonizzoni2017coherent}%
  \BibitemOpen
  \bibfield  {author} {\bibinfo {author} {\bibfnamefont {C.}~\bibnamefont {Bonizzoni}}, \bibinfo {author} {\bibfnamefont {A.}~\bibnamefont {Ghirri}}, \bibinfo {author} {\bibfnamefont {M.}~\bibnamefont {Atzori}}, \bibinfo {author} {\bibfnamefont {L.}~\bibnamefont {Sorace}}, \bibinfo {author} {\bibfnamefont {R.}~\bibnamefont {Sessoli}},\ and\ \bibinfo {author} {\bibfnamefont {M.}~\bibnamefont {Affronte}},\ }\bibfield  {title} {\bibinfo {title} {Coherent coupling between vanadyl phthalocyanine spin ensemble and microwave photons: towards integration of molecular spin qubits into quantum circuits},\ }\href@noop {} {\bibfield  {journal} {\bibinfo  {journal} {Scientific {R}eports}\ }\textbf {\bibinfo {volume} {7}},\ \bibinfo {pages} {13096} (\bibinfo {year} {2017})}\BibitemShut {NoStop}%
\bibitem [{\citenamefont {Malavolti}\ \emph {et~al.}(2018)\citenamefont {Malavolti}, \citenamefont {Briganti}, \citenamefont {H{\"a}nze}, \citenamefont {Serrano}, \citenamefont {Cimatti}, \citenamefont {McMurtrie}, \citenamefont {Otero}, \citenamefont {Ohresser}, \citenamefont {Totti}, \citenamefont {Mannini}, \citenamefont {Sessoli},\ and\ \citenamefont {Loth}}]{malavolti2018tunable}%
  \BibitemOpen
  \bibfield  {author} {\bibinfo {author} {\bibfnamefont {L.}~\bibnamefont {Malavolti}}, \bibinfo {author} {\bibfnamefont {M.}~\bibnamefont {Briganti}}, \bibinfo {author} {\bibfnamefont {M.}~\bibnamefont {H{\"a}nze}}, \bibinfo {author} {\bibfnamefont {G.}~\bibnamefont {Serrano}}, \bibinfo {author} {\bibfnamefont {I.}~\bibnamefont {Cimatti}}, \bibinfo {author} {\bibfnamefont {G.}~\bibnamefont {McMurtrie}}, \bibinfo {author} {\bibfnamefont {E.}~\bibnamefont {Otero}}, \bibinfo {author} {\bibfnamefont {P.}~\bibnamefont {Ohresser}}, \bibinfo {author} {\bibfnamefont {F.}~\bibnamefont {Totti}}, \bibinfo {author} {\bibfnamefont {M.}~\bibnamefont {Mannini}}, \bibinfo {author} {\bibfnamefont {R.}~\bibnamefont {Sessoli}},\ and\ \bibinfo {author} {\bibfnamefont {S.}~\bibnamefont {Loth}},\ }\bibfield  {title} {\bibinfo {title} {Tunable spin--superconductor coupling of spin 1/2 vanadyl phthalocyanine molecules},\ }\href {https://doi.org/10.1021/acs.nanolett.8b03921} {\bibfield  {journal} {\bibinfo  {journal} {Nano
  {L}etters}\ }\textbf {\bibinfo {volume} {18}},\ \bibinfo {pages} {7955} (\bibinfo {year} {2018})}\BibitemShut {NoStop}%
\bibitem [{\citenamefont {Bader}\ \emph {et~al.}(2016)\citenamefont {Bader}, \citenamefont {Winkler},\ and\ \citenamefont {van Slageren}}]{bader2016tuning}%
  \BibitemOpen
  \bibfield  {author} {\bibinfo {author} {\bibfnamefont {K.}~\bibnamefont {Bader}}, \bibinfo {author} {\bibfnamefont {M.}~\bibnamefont {Winkler}},\ and\ \bibinfo {author} {\bibfnamefont {J.}~\bibnamefont {van Slageren}},\ }\bibfield  {title} {\bibinfo {title} {Tuning of molecular qubits: very long coherence and spin--lattice relaxation times},\ }\href@noop {} {\bibfield  {journal} {\bibinfo  {journal} {Chemical {C}ommunications}\ }\textbf {\bibinfo {volume} {52}},\ \bibinfo {pages} {3623} (\bibinfo {year} {2016})}\BibitemShut {NoStop}%
\bibitem [{\citenamefont {Cimatti}\ \emph {et~al.}(2019)\citenamefont {Cimatti}, \citenamefont {Bond{\`\i}}, \citenamefont {Serrano}, \citenamefont {Malavolti}, \citenamefont {Cortigiani}, \citenamefont {Velez-Fort}, \citenamefont {Betto}, \citenamefont {Ouerghi}, \citenamefont {Brookes}, \citenamefont {Loth} \emph {et~al.}}]{cimatti2019vanadyl}%
  \BibitemOpen
  \bibfield  {author} {\bibinfo {author} {\bibfnamefont {I.}~\bibnamefont {Cimatti}}, \bibinfo {author} {\bibfnamefont {L.}~\bibnamefont {Bond{\`\i}}}, \bibinfo {author} {\bibfnamefont {G.}~\bibnamefont {Serrano}}, \bibinfo {author} {\bibfnamefont {L.}~\bibnamefont {Malavolti}}, \bibinfo {author} {\bibfnamefont {B.}~\bibnamefont {Cortigiani}}, \bibinfo {author} {\bibfnamefont {E.}~\bibnamefont {Velez-Fort}}, \bibinfo {author} {\bibfnamefont {D.}~\bibnamefont {Betto}}, \bibinfo {author} {\bibfnamefont {A.}~\bibnamefont {Ouerghi}}, \bibinfo {author} {\bibfnamefont {N.}~\bibnamefont {Brookes}}, \bibinfo {author} {\bibfnamefont {S.}~\bibnamefont {Loth}}, \emph {et~al.},\ }\bibfield  {title} {\bibinfo {title} {Vanadyl phthalocyanines on graphene/{SiC} (0001): toward a hybrid architecture for molecular spin qubits},\ }\href@noop {} {\bibfield  {journal} {\bibinfo  {journal} {Nanoscale {H}orizons}\ }\textbf {\bibinfo {volume} {4}},\ \bibinfo {pages} {1202} (\bibinfo {year} {2019})}\BibitemShut {NoStop}%
\bibitem [{\citenamefont {Burkard}\ \emph {et~al.}(2023)\citenamefont {Burkard}, \citenamefont {Ladd}, \citenamefont {Pan}, \citenamefont {Nichol},\ and\ \citenamefont {Petta}}]{burkard2023semiconductor}%
  \BibitemOpen
  \bibfield  {author} {\bibinfo {author} {\bibfnamefont {G.}~\bibnamefont {Burkard}}, \bibinfo {author} {\bibfnamefont {T.~D.}\ \bibnamefont {Ladd}}, \bibinfo {author} {\bibfnamefont {A.}~\bibnamefont {Pan}}, \bibinfo {author} {\bibfnamefont {J.~M.}\ \bibnamefont {Nichol}},\ and\ \bibinfo {author} {\bibfnamefont {J.~R.}\ \bibnamefont {Petta}},\ }\bibfield  {title} {\bibinfo {title} {Semiconductor spin qubits},\ }\href@noop {} {\bibfield  {journal} {\bibinfo  {journal} {Reviews of {M}odern {P}hysics}\ }\textbf {\bibinfo {volume} {95}},\ \bibinfo {pages} {025003} (\bibinfo {year} {2023})}\BibitemShut {NoStop}%
\bibitem [{\citenamefont {Noh}\ \emph {et~al.}(2023{\natexlab{a}})\citenamefont {Noh}, \citenamefont {Colazzo}, \citenamefont {Urdaniz}, \citenamefont {Lee}, \citenamefont {Krylov}, \citenamefont {Devi}, \citenamefont {Doll}, \citenamefont {Heinrich}, \citenamefont {Wolf}, \citenamefont {Donati} \emph {et~al.}}]{noh2023vanadyl}%
  \BibitemOpen
  \bibfield  {author} {\bibinfo {author} {\bibfnamefont {K.}~\bibnamefont {Noh}}, \bibinfo {author} {\bibfnamefont {L.}~\bibnamefont {Colazzo}}, \bibinfo {author} {\bibfnamefont {C.}~\bibnamefont {Urdaniz}}, \bibinfo {author} {\bibfnamefont {J.}~\bibnamefont {Lee}}, \bibinfo {author} {\bibfnamefont {D.}~\bibnamefont {Krylov}}, \bibinfo {author} {\bibfnamefont {P.}~\bibnamefont {Devi}}, \bibinfo {author} {\bibfnamefont {A.}~\bibnamefont {Doll}}, \bibinfo {author} {\bibfnamefont {A.}~\bibnamefont {Heinrich}}, \bibinfo {author} {\bibfnamefont {C.}~\bibnamefont {Wolf}}, \bibinfo {author} {\bibfnamefont {F.}~\bibnamefont {Donati}}, \emph {et~al.},\ }\bibfield  {title} {\bibinfo {title} {Vanadyl phthalocyanine single molecules on insulating layers; ultrathin mgo film and titanyl phthalocyanine monolayer},\ }in\ \href@noop {} {\emph {\bibinfo {booktitle} {{APS} March Meeting Abstracts}}},\ Vol.\ \bibinfo {volume} {2023}\ (\bibinfo {year} {2023})\ pp.\ \bibinfo {pages} {G58--004}\BibitemShut {NoStop}%
\bibitem [{\citenamefont {Xu}\ \emph {et~al.}(2022)\citenamefont {Xu}, \citenamefont {Romankov}, \citenamefont {Doll},\ and\ \citenamefont {Dreiser}}]{xu2022orienting}%
  \BibitemOpen
  \bibfield  {author} {\bibinfo {author} {\bibfnamefont {Z.}~\bibnamefont {Xu}}, \bibinfo {author} {\bibfnamefont {V.}~\bibnamefont {Romankov}}, \bibinfo {author} {\bibfnamefont {A.}~\bibnamefont {Doll}},\ and\ \bibinfo {author} {\bibfnamefont {J.}~\bibnamefont {Dreiser}},\ }\bibfield  {title} {\bibinfo {title} {Orienting dilute thin films of non-planar spin-1/2 vanadyl--phthalocyanine complexes},\ }\href@noop {} {\bibfield  {journal} {\bibinfo  {journal} {Materials {A}dvances}\ }\textbf {\bibinfo {volume} {3}},\ \bibinfo {pages} {4938} (\bibinfo {year} {2022})}\BibitemShut {NoStop}%
\bibitem [{\citenamefont {Lee}\ \emph {et~al.}(2024)\citenamefont {Lee}, \citenamefont {Urdaniz}, \citenamefont {Reale}, \citenamefont {Noh}, \citenamefont {Krylov}, \citenamefont {Doll}, \citenamefont {Colazzo}, \citenamefont {Bae}, \citenamefont {Wolf},\ and\ \citenamefont {Donati}}]{lee2024interpreting}%
  \BibitemOpen
  \bibfield  {author} {\bibinfo {author} {\bibfnamefont {J.}~\bibnamefont {Lee}}, \bibinfo {author} {\bibfnamefont {C.}~\bibnamefont {Urdaniz}}, \bibinfo {author} {\bibfnamefont {S.}~\bibnamefont {Reale}}, \bibinfo {author} {\bibfnamefont {K.}~\bibnamefont {Noh}}, \bibinfo {author} {\bibfnamefont {D.}~\bibnamefont {Krylov}}, \bibinfo {author} {\bibfnamefont {A.}~\bibnamefont {Doll}}, \bibinfo {author} {\bibfnamefont {L.}~\bibnamefont {Colazzo}}, \bibinfo {author} {\bibfnamefont {Y.}~\bibnamefont {Bae}}, \bibinfo {author} {\bibfnamefont {C.}~\bibnamefont {Wolf}},\ and\ \bibinfo {author} {\bibfnamefont {F.}~\bibnamefont {Donati}},\ }\bibfield  {title} {\bibinfo {title} {Interpreting x-ray absorption spectra of vanadyl phthalocyanines spin qubit candidates using a machine learning assisted approach},\ }\href@noop {} {\bibfield  {journal} {\bibinfo  {journal} {Physical {R}eview {B}}\ }\textbf {\bibinfo {volume} {109}},\ \bibinfo {pages} {235427} (\bibinfo {year} {2024})}\BibitemShut {NoStop}%
\bibitem [{\citenamefont {Noh}\ \emph {et~al.}(2023{\natexlab{b}})\citenamefont {Noh}, \citenamefont {Colazzo}, \citenamefont {Urdaniz}, \citenamefont {Lee}, \citenamefont {Krylov}, \citenamefont {Devi}, \citenamefont {Doll}, \citenamefont {Heinrich}, \citenamefont {Wolf}, \citenamefont {Donati} \emph {et~al.}}]{noh2023template}%
  \BibitemOpen
  \bibfield  {author} {\bibinfo {author} {\bibfnamefont {K.}~\bibnamefont {Noh}}, \bibinfo {author} {\bibfnamefont {L.}~\bibnamefont {Colazzo}}, \bibinfo {author} {\bibfnamefont {C.}~\bibnamefont {Urdaniz}}, \bibinfo {author} {\bibfnamefont {J.}~\bibnamefont {Lee}}, \bibinfo {author} {\bibfnamefont {D.}~\bibnamefont {Krylov}}, \bibinfo {author} {\bibfnamefont {P.}~\bibnamefont {Devi}}, \bibinfo {author} {\bibfnamefont {A.}~\bibnamefont {Doll}}, \bibinfo {author} {\bibfnamefont {A.~J.}\ \bibnamefont {Heinrich}}, \bibinfo {author} {\bibfnamefont {C.}~\bibnamefont {Wolf}}, \bibinfo {author} {\bibfnamefont {F.}~\bibnamefont {Donati}}, \emph {et~al.},\ }\bibfield  {title} {\bibinfo {title} {Template-directed {2D} nanopatterning of {S}= 1/2 molecular spins},\ }\href@noop {} {\bibfield  {journal} {\bibinfo  {journal} {Nanoscale {H}orizons}\ }\textbf {\bibinfo {volume} {8}},\ \bibinfo {pages} {624} (\bibinfo {year} {2023}{\natexlab{b}})}\BibitemShut {NoStop}%
\bibitem [{\citenamefont {Sun}\ \emph {et~al.}(2018)\citenamefont {Sun}, \citenamefont {Kim}, \citenamefont {Solomon},\ and\ \citenamefont {Waks}}]{sun2018cavity}%
  \BibitemOpen
  \bibfield  {author} {\bibinfo {author} {\bibfnamefont {S.}~\bibnamefont {Sun}}, \bibinfo {author} {\bibfnamefont {H.}~\bibnamefont {Kim}}, \bibinfo {author} {\bibfnamefont {G.~S.}\ \bibnamefont {Solomon}},\ and\ \bibinfo {author} {\bibfnamefont {E.}~\bibnamefont {Waks}},\ }\bibfield  {title} {\bibinfo {title} {Cavity-enhanced optical readout of a single solid-state spin},\ }\href@noop {} {\bibfield  {journal} {\bibinfo  {journal} {Physical {R}eview {A}pplied}\ }\textbf {\bibinfo {volume} {9}},\ \bibinfo {pages} {054013} (\bibinfo {year} {2018})}\BibitemShut {NoStop}%
\bibitem [{\citenamefont {Collins}\ \emph {et~al.}(1993)\citenamefont {Collins}, \citenamefont {Krier},\ and\ \citenamefont {Abass}}]{collins1993optical}%
  \BibitemOpen
  \bibfield  {author} {\bibinfo {author} {\bibfnamefont {R.}~\bibnamefont {Collins}}, \bibinfo {author} {\bibfnamefont {A.}~\bibnamefont {Krier}},\ and\ \bibinfo {author} {\bibfnamefont {A.}~\bibnamefont {Abass}},\ }\bibfield  {title} {\bibinfo {title} {Optical properties of lead phthalocyanine ({PbPc}) thin films},\ }\href@noop {} {\bibfield  {journal} {\bibinfo  {journal} {Thin {S}olid {F}ilms}\ }\textbf {\bibinfo {volume} {229}},\ \bibinfo {pages} {113} (\bibinfo {year} {1993})}\BibitemShut {NoStop}%
\bibitem [{\citenamefont {Klyamer}\ \emph {et~al.}(2020)\citenamefont {Klyamer}, \citenamefont {Sukhikh}, \citenamefont {Nikolaeva}, \citenamefont {Morozova},\ and\ \citenamefont {Basova}}]{klyamer2020vanadyl}%
  \BibitemOpen
  \bibfield  {author} {\bibinfo {author} {\bibfnamefont {D.}~\bibnamefont {Klyamer}}, \bibinfo {author} {\bibfnamefont {A.}~\bibnamefont {Sukhikh}}, \bibinfo {author} {\bibfnamefont {N.}~\bibnamefont {Nikolaeva}}, \bibinfo {author} {\bibfnamefont {N.}~\bibnamefont {Morozova}},\ and\ \bibinfo {author} {\bibfnamefont {T.}~\bibnamefont {Basova}},\ }\bibfield  {title} {\bibinfo {title} {Vanadyl phthalocyanine films and their hybrid structures with {Pd} nanoparticles: Structure and sensing properties},\ }\href@noop {} {\bibfield  {journal} {\bibinfo  {journal} {Sensors}\ }\textbf {\bibinfo {volume} {20}},\ \bibinfo {pages} {1893} (\bibinfo {year} {2020})}\BibitemShut {NoStop}%
\bibitem [{\citenamefont {Wang}\ \emph {et~al.}(2014)\citenamefont {Wang}, \citenamefont {Zhou}, \citenamefont {Roy}, \citenamefont {Yan}, \citenamefont {Zhang},\ and\ \citenamefont {Lee}}]{wang2014polymorphism}%
  \BibitemOpen
  \bibfield  {author} {\bibinfo {author} {\bibfnamefont {H.}~\bibnamefont {Wang}}, \bibinfo {author} {\bibfnamefont {Y.}~\bibnamefont {Zhou}}, \bibinfo {author} {\bibfnamefont {V.}~\bibnamefont {Roy}}, \bibinfo {author} {\bibfnamefont {D.}~\bibnamefont {Yan}}, \bibinfo {author} {\bibfnamefont {J.}~\bibnamefont {Zhang}},\ and\ \bibinfo {author} {\bibfnamefont {C.-S.}\ \bibnamefont {Lee}},\ }\bibfield  {title} {\bibinfo {title} {Polymorphism and electronic properties of vanadyl-phthalocyanine films},\ }\href@noop {} {\bibfield  {journal} {\bibinfo  {journal} {Organic {E}lectronics}\ }\textbf {\bibinfo {volume} {15}},\ \bibinfo {pages} {1586} (\bibinfo {year} {2014})}\BibitemShut {NoStop}%
\bibitem [{\citenamefont {Fang}\ \emph {et~al.}(1996)\citenamefont {Fang}, \citenamefont {Hoshi}, \citenamefont {Kohama},\ and\ \citenamefont {Maruyama}}]{fang1996nonlinear}%
  \BibitemOpen
  \bibfield  {author} {\bibinfo {author} {\bibfnamefont {S.}~\bibnamefont {Fang}}, \bibinfo {author} {\bibfnamefont {H.}~\bibnamefont {Hoshi}}, \bibinfo {author} {\bibfnamefont {K.}~\bibnamefont {Kohama}},\ and\ \bibinfo {author} {\bibfnamefont {Y.}~\bibnamefont {Maruyama}},\ }\bibfield  {title} {\bibinfo {title} {Nonlinear optical characteristics of vanadyl phthalocyanine thin film grown by the molecular beam epitaxial method},\ }\href@noop {} {\bibfield  {journal} {\bibinfo  {journal} {The {J}ournal of {P}hysical {C}hemistry}\ }\textbf {\bibinfo {volume} {100}},\ \bibinfo {pages} {4104} (\bibinfo {year} {1996})}\BibitemShut {NoStop}%
\bibitem [{\citenamefont {Papageorgiou}\ \emph {et~al.}(2004)\citenamefont {Papageorgiou}, \citenamefont {Salomon}, \citenamefont {Angot}, \citenamefont {Layet}, \citenamefont {Giovanelli},\ and\ \citenamefont {Le~Lay}}]{papageorgiou2004physics}%
  \BibitemOpen
  \bibfield  {author} {\bibinfo {author} {\bibfnamefont {N.}~\bibnamefont {Papageorgiou}}, \bibinfo {author} {\bibfnamefont {E.}~\bibnamefont {Salomon}}, \bibinfo {author} {\bibfnamefont {T.}~\bibnamefont {Angot}}, \bibinfo {author} {\bibfnamefont {J.-M.}\ \bibnamefont {Layet}}, \bibinfo {author} {\bibfnamefont {L.}~\bibnamefont {Giovanelli}},\ and\ \bibinfo {author} {\bibfnamefont {G.}~\bibnamefont {Le~Lay}},\ }\bibfield  {title} {\bibinfo {title} {Physics of ultra-thin phthalocyanine films on semiconductors},\ }\href@noop {} {\bibfield  {journal} {\bibinfo  {journal} {Progress in {S}urface {S}cience}\ }\textbf {\bibinfo {volume} {77}},\ \bibinfo {pages} {139} (\bibinfo {year} {2004})}\BibitemShut {NoStop}%
\bibitem [{\citenamefont {Yu}\ \emph {et~al.}(2007)\citenamefont {Yu}, \citenamefont {Xu}, \citenamefont {Cheung},\ and\ \citenamefont {Ke}}]{yu2007optimizing}%
  \BibitemOpen
  \bibfield  {author} {\bibinfo {author} {\bibfnamefont {X.~J.}\ \bibnamefont {Yu}}, \bibinfo {author} {\bibfnamefont {J.~B.}\ \bibnamefont {Xu}}, \bibinfo {author} {\bibfnamefont {W.~Y.}\ \bibnamefont {Cheung}},\ and\ \bibinfo {author} {\bibfnamefont {N.}~\bibnamefont {Ke}},\ }\bibfield  {title} {\bibinfo {title} {Optimizing the growth of vanadyl-phthalocyanine thin films for high-mobility organic thin-film transistors},\ }\href@noop {} {\bibfield  {journal} {\bibinfo  {journal} {Journal of {A}pplied {P}hysics}\ }\textbf {\bibinfo {volume} {102}},\ \bibinfo {pages} {103711/1} (\bibinfo {year} {2007})}\BibitemShut {NoStop}%
\bibitem [{\citenamefont {Pakhomov}\ \emph {et~al.}(2010)\citenamefont {Pakhomov}, \citenamefont {Pakhomov}, \citenamefont {Travkin}, \citenamefont {Abanin}, \citenamefont {Stakhira},\ and\ \citenamefont {Cherpak}}]{pakhomov2010phthalocyanine}%
  \BibitemOpen
  \bibfield  {author} {\bibinfo {author} {\bibfnamefont {G.~L.}\ \bibnamefont {Pakhomov}}, \bibinfo {author} {\bibfnamefont {L.~G.}\ \bibnamefont {Pakhomov}}, \bibinfo {author} {\bibfnamefont {V.~V.}\ \bibnamefont {Travkin}}, \bibinfo {author} {\bibfnamefont {M.~V.}\ \bibnamefont {Abanin}}, \bibinfo {author} {\bibfnamefont {P.~Y.}\ \bibnamefont {Stakhira}},\ and\ \bibinfo {author} {\bibfnamefont {V.~V.}\ \bibnamefont {Cherpak}},\ }\bibfield  {title} {\bibinfo {title} {Phthalocyanine-based photoelectrical cells: effect of environment on power conversion efficiency},\ }\href@noop {} {\bibfield  {journal} {\bibinfo  {journal} {Journal of {M}aterials {S}cience}\ }\textbf {\bibinfo {volume} {45}},\ \bibinfo {pages} {1854} (\bibinfo {year} {2010})}\BibitemShut {NoStop}%
\bibitem [{\citenamefont {Eguchi}\ \emph {et~al.}(2013)\citenamefont {Eguchi}, \citenamefont {Takagi}, \citenamefont {Nakagawa},\ and\ \citenamefont {Yokoyama}}]{eguchi2013molecular}%
  \BibitemOpen
  \bibfield  {author} {\bibinfo {author} {\bibfnamefont {K.}~\bibnamefont {Eguchi}}, \bibinfo {author} {\bibfnamefont {Y.}~\bibnamefont {Takagi}}, \bibinfo {author} {\bibfnamefont {T.}~\bibnamefont {Nakagawa}},\ and\ \bibinfo {author} {\bibfnamefont {T.}~\bibnamefont {Yokoyama}},\ }\bibfield  {title} {\bibinfo {title} {Molecular orientation and electronic states of vanadyl phthalocyanine on {S}i (111) and {A}g (111) surfaces},\ }\href@noop {} {\bibfield  {journal} {\bibinfo  {journal} {The {J}ournal of {P}hysical {C}hemistry {C}}\ }\textbf {\bibinfo {volume} {117}},\ \bibinfo {pages} {22843} (\bibinfo {year} {2013})}\BibitemShut {NoStop}%
\bibitem [{\citenamefont {Amsterdam}\ \emph {et~al.}(2019)\citenamefont {Amsterdam}, \citenamefont {Stanev}, \citenamefont {Zhou}, \citenamefont {Lou}, \citenamefont {Bergeron}, \citenamefont {Darancet}, \citenamefont {Hersam}, \citenamefont {Stern},\ and\ \citenamefont {Marks}}]{amsterdam2019electronic}%
  \BibitemOpen
  \bibfield  {author} {\bibinfo {author} {\bibfnamefont {S.~H.}\ \bibnamefont {Amsterdam}}, \bibinfo {author} {\bibfnamefont {T.~K.}\ \bibnamefont {Stanev}}, \bibinfo {author} {\bibfnamefont {Q.}~\bibnamefont {Zhou}}, \bibinfo {author} {\bibfnamefont {A.~J.-T.}\ \bibnamefont {Lou}}, \bibinfo {author} {\bibfnamefont {H.}~\bibnamefont {Bergeron}}, \bibinfo {author} {\bibfnamefont {P.}~\bibnamefont {Darancet}}, \bibinfo {author} {\bibfnamefont {M.~C.}\ \bibnamefont {Hersam}}, \bibinfo {author} {\bibfnamefont {N.~P.}\ \bibnamefont {Stern}},\ and\ \bibinfo {author} {\bibfnamefont {T.~J.}\ \bibnamefont {Marks}},\ }\bibfield  {title} {\bibinfo {title} {Electronic coupling in metallophthalocyanine--transition metal dichalcogenide mixed-dimensional heterojunctions},\ }\href@noop {} {\bibfield  {journal} {\bibinfo  {journal} {{ACS} {N}ano}\ }\textbf {\bibinfo {volume} {13}},\ \bibinfo {pages} {4183} (\bibinfo {year} {2019})}\BibitemShut {NoStop}%
\bibitem [{\citenamefont {Blowey}\ \emph {et~al.}(2018)\citenamefont {Blowey}, \citenamefont {Maurer}, \citenamefont {Rochford}, \citenamefont {Duncan}, \citenamefont {Kang}, \citenamefont {Warr}, \citenamefont {Ramadan}, \citenamefont {Lee}, \citenamefont {Thakur}, \citenamefont {Costantini} \emph {et~al.}}]{blowey2018structure}%
  \BibitemOpen
  \bibfield  {author} {\bibinfo {author} {\bibfnamefont {P.~J.}\ \bibnamefont {Blowey}}, \bibinfo {author} {\bibfnamefont {R.~J.}\ \bibnamefont {Maurer}}, \bibinfo {author} {\bibfnamefont {L.~A.}\ \bibnamefont {Rochford}}, \bibinfo {author} {\bibfnamefont {D.}~\bibnamefont {Duncan}}, \bibinfo {author} {\bibfnamefont {J.-H.}\ \bibnamefont {Kang}}, \bibinfo {author} {\bibfnamefont {D.}~\bibnamefont {Warr}}, \bibinfo {author} {\bibfnamefont {A.}~\bibnamefont {Ramadan}}, \bibinfo {author} {\bibfnamefont {T.-L.}\ \bibnamefont {Lee}}, \bibinfo {author} {\bibfnamefont {P.}~\bibnamefont {Thakur}}, \bibinfo {author} {\bibfnamefont {G.}~\bibnamefont {Costantini}}, \emph {et~al.},\ }\bibfield  {title} {\bibinfo {title} {The structure of {VOPc} on {C}u (111): Does {V=O} point up, or down, or both?},\ }\href@noop {} {\bibfield  {journal} {\bibinfo  {journal} {Journal of {P}hysical {C}hemistry {C}}\ }\textbf {\bibinfo {volume} {123}},\ \bibinfo {pages} {8101} (\bibinfo {year} {2018})}\BibitemShut {NoStop}%
\bibitem [{\citenamefont {Niu}\ \emph {et~al.}(2014)\citenamefont {Niu}, \citenamefont {Zhang},\ and\ \citenamefont {Chen}}]{niu2014molecular}%
  \BibitemOpen
  \bibfield  {author} {\bibinfo {author} {\bibfnamefont {T.}~\bibnamefont {Niu}}, \bibinfo {author} {\bibfnamefont {J.}~\bibnamefont {Zhang}},\ and\ \bibinfo {author} {\bibfnamefont {W.}~\bibnamefont {Chen}},\ }\bibfield  {title} {\bibinfo {title} {Molecular ordering and dipole alignment of vanadyl phthalocyanine monolayer on metals: The effects of interfacial interactions},\ }\href@noop {} {\bibfield  {journal} {\bibinfo  {journal} {The {J}ournal of {P}hysical {C}hemistry {C}}\ }\textbf {\bibinfo {volume} {118}},\ \bibinfo {pages} {4151} (\bibinfo {year} {2014})}\BibitemShut {NoStop}%
\bibitem [{\citenamefont {Griffiths}\ \emph {et~al.}(1976)\citenamefont {Griffiths}, \citenamefont {Walker},\ and\ \citenamefont {Goldstein}}]{griffiths1976polymorphism}%
  \BibitemOpen
  \bibfield  {author} {\bibinfo {author} {\bibfnamefont {C.~H.}\ \bibnamefont {Griffiths}}, \bibinfo {author} {\bibfnamefont {M.~S.}\ \bibnamefont {Walker}},\ and\ \bibinfo {author} {\bibfnamefont {P.}~\bibnamefont {Goldstein}},\ }\bibfield  {title} {\bibinfo {title} {Polymorphism in vanadyl phthalocyanine},\ }\href {https://doi.org/10.1080/15421407608083878} {\bibfield  {journal} {\bibinfo  {journal} {Molecular {C}rystals and {L}iquid {C}rystals}\ }\textbf {\bibinfo {volume} {33}},\ \bibinfo {pages} {149} (\bibinfo {year} {1976})}\BibitemShut {NoStop}%
\bibitem [{\citenamefont {Ziolo}\ \emph {et~al.}(1980)\citenamefont {Ziolo}, \citenamefont {Griffiths},\ and\ \citenamefont {Troup}}]{ziolo1980phaseii}%
  \BibitemOpen
  \bibfield  {author} {\bibinfo {author} {\bibfnamefont {R.~F.}\ \bibnamefont {Ziolo}}, \bibinfo {author} {\bibfnamefont {C.~H.}\ \bibnamefont {Griffiths}},\ and\ \bibinfo {author} {\bibfnamefont {J.~M.}\ \bibnamefont {Troup}},\ }\bibfield  {title} {\bibinfo {title} {Crystal structure of vanadyl phthalocyanine, phase ii},\ }\href {https://doi.org/10.1039/DT9800002300} {\bibfield  {journal} {\bibinfo  {journal} {Journal of the Chemical Society, Dalton Transactions}\ ,\ \bibinfo {pages} {2300}} (\bibinfo {year} {1980})}\BibitemShut {NoStop}%
\bibitem [{\citenamefont {Ramadan}\ \emph {et~al.}(2016)\citenamefont {Ramadan}, \citenamefont {Rochford}, \citenamefont {Moffat}, \citenamefont {Mulcahy}, \citenamefont {Ryan}, \citenamefont {Jones},\ and\ \citenamefont {Heutz}}]{ramadan2016morphology}%
  \BibitemOpen
  \bibfield  {author} {\bibinfo {author} {\bibfnamefont {A.~J.}\ \bibnamefont {Ramadan}}, \bibinfo {author} {\bibfnamefont {L.~A.}\ \bibnamefont {Rochford}}, \bibinfo {author} {\bibfnamefont {J.}~\bibnamefont {Moffat}}, \bibinfo {author} {\bibfnamefont {C.}~\bibnamefont {Mulcahy}}, \bibinfo {author} {\bibfnamefont {M.~P.}\ \bibnamefont {Ryan}}, \bibinfo {author} {\bibfnamefont {T.~S.}\ \bibnamefont {Jones}},\ and\ \bibinfo {author} {\bibfnamefont {S.}~\bibnamefont {Heutz}},\ }\bibfield  {title} {\bibinfo {title} {The morphology and structure of vanadyl phthalocyanine thin films on lithium niobate single crystals},\ }\href@noop {} {\bibfield  {journal} {\bibinfo  {journal} {Journal of {M}aterials {C}hemistry {C}}\ }\textbf {\bibinfo {volume} {4}},\ \bibinfo {pages} {348} (\bibinfo {year} {2016})}\BibitemShut {NoStop}%
\bibitem [{\citenamefont {Amsterdam}\ \emph {et~al.}(2020)\citenamefont {Amsterdam}, \citenamefont {LaMountain}, \citenamefont {Stanev}, \citenamefont {Sangwan}, \citenamefont {L{\'o}pez-Arteaga}, \citenamefont {Padgaonkar}, \citenamefont {Watanabe}, \citenamefont {Taniguchi}, \citenamefont {Weiss}, \citenamefont {Marks} \emph {et~al.}}]{amsterdam2020tailoring}%
  \BibitemOpen
  \bibfield  {author} {\bibinfo {author} {\bibfnamefont {S.~H.}\ \bibnamefont {Amsterdam}}, \bibinfo {author} {\bibfnamefont {T.}~\bibnamefont {LaMountain}}, \bibinfo {author} {\bibfnamefont {T.~K.}\ \bibnamefont {Stanev}}, \bibinfo {author} {\bibfnamefont {V.~K.}\ \bibnamefont {Sangwan}}, \bibinfo {author} {\bibfnamefont {R.}~\bibnamefont {L{\'o}pez-Arteaga}}, \bibinfo {author} {\bibfnamefont {S.}~\bibnamefont {Padgaonkar}}, \bibinfo {author} {\bibfnamefont {K.}~\bibnamefont {Watanabe}}, \bibinfo {author} {\bibfnamefont {T.}~\bibnamefont {Taniguchi}}, \bibinfo {author} {\bibfnamefont {E.~A.}\ \bibnamefont {Weiss}}, \bibinfo {author} {\bibfnamefont {T.~J.}\ \bibnamefont {Marks}}, \emph {et~al.},\ }\bibfield  {title} {\bibinfo {title} {Tailoring the optical response of pentacene thin films via templated growth on hexagonal boron nitride},\ }\href@noop {} {\bibfield  {journal} {\bibinfo  {journal} {Journal of {P}hysical {C}hemistry {L}etters}\ }\textbf {\bibinfo {volume} {12}},\ \bibinfo {pages} {26} (\bibinfo
  {year} {2020})}\BibitemShut {NoStop}%
\bibitem [{\citenamefont {Geim}\ and\ \citenamefont {Grigorieva}(2013)}]{geim2013van}%
  \BibitemOpen
  \bibfield  {author} {\bibinfo {author} {\bibfnamefont {A.~K.}\ \bibnamefont {Geim}}\ and\ \bibinfo {author} {\bibfnamefont {I.~V.}\ \bibnamefont {Grigorieva}},\ }\bibfield  {title} {\bibinfo {title} {Van der {W}aals heterostructures},\ }\href@noop {} {\bibfield  {journal} {\bibinfo  {journal} {Nature}\ }\textbf {\bibinfo {volume} {499}},\ \bibinfo {pages} {419} (\bibinfo {year} {2013})}\BibitemShut {NoStop}%
\bibitem [{\citenamefont {Blackstone}\ and\ \citenamefont {Ignaszak}(2021)}]{blackstone2021van}%
  \BibitemOpen
  \bibfield  {author} {\bibinfo {author} {\bibfnamefont {C.}~\bibnamefont {Blackstone}}\ and\ \bibinfo {author} {\bibfnamefont {A.}~\bibnamefont {Ignaszak}},\ }\bibfield  {title} {\bibinfo {title} {Van der {W}aals heterostructures—{R}ecent progress in electrode materials for clean energy applications},\ }\href@noop {} {\bibfield  {journal} {\bibinfo  {journal} {Materials}\ }\textbf {\bibinfo {volume} {14}},\ \bibinfo {pages} {3754} (\bibinfo {year} {2021})}\BibitemShut {NoStop}%
\bibitem [{\citenamefont {Novoselov}\ \emph {et~al.}(2016)\citenamefont {Novoselov}, \citenamefont {Mishchenko}, \citenamefont {Carvalho},\ and\ \citenamefont {Castro~Neto}}]{novoselov20162d}%
  \BibitemOpen
  \bibfield  {author} {\bibinfo {author} {\bibfnamefont {K.~S.}\ \bibnamefont {Novoselov}}, \bibinfo {author} {\bibfnamefont {A.}~\bibnamefont {Mishchenko}}, \bibinfo {author} {\bibfnamefont {A.}~\bibnamefont {Carvalho}},\ and\ \bibinfo {author} {\bibfnamefont {A.}~\bibnamefont {Castro~Neto}},\ }\bibfield  {title} {\bibinfo {title} {2d materials and van der {W}aals heterostructures},\ }\href@noop {} {\bibfield  {journal} {\bibinfo  {journal} {Science}\ }\textbf {\bibinfo {volume} {353}},\ \bibinfo {pages} {aac9439} (\bibinfo {year} {2016})}\BibitemShut {NoStop}%
\bibitem [{\citenamefont {Jariwala}\ \emph {et~al.}(2017)\citenamefont {Jariwala}, \citenamefont {Marks},\ and\ \citenamefont {Hersam}}]{jariwala2017mixed}%
  \BibitemOpen
  \bibfield  {author} {\bibinfo {author} {\bibfnamefont {D.}~\bibnamefont {Jariwala}}, \bibinfo {author} {\bibfnamefont {T.~J.}\ \bibnamefont {Marks}},\ and\ \bibinfo {author} {\bibfnamefont {M.~C.}\ \bibnamefont {Hersam}},\ }\bibfield  {title} {\bibinfo {title} {Mixed-dimensional van der {W}aals heterostructures},\ }\href@noop {} {\bibfield  {journal} {\bibinfo  {journal} {Nature {M}aterials}\ }\textbf {\bibinfo {volume} {16}},\ \bibinfo {pages} {170} (\bibinfo {year} {2017})}\BibitemShut {NoStop}%
\bibitem [{\citenamefont {Padgaonkar}\ \emph {et~al.}(2019)\citenamefont {Padgaonkar}, \citenamefont {Amsterdam}, \citenamefont {Bergeron}, \citenamefont {Su}, \citenamefont {Marks}, \citenamefont {Hersam},\ and\ \citenamefont {Weiss}}]{padgaonkar2019molecular}%
  \BibitemOpen
  \bibfield  {author} {\bibinfo {author} {\bibfnamefont {S.}~\bibnamefont {Padgaonkar}}, \bibinfo {author} {\bibfnamefont {S.~H.}\ \bibnamefont {Amsterdam}}, \bibinfo {author} {\bibfnamefont {H.}~\bibnamefont {Bergeron}}, \bibinfo {author} {\bibfnamefont {K.}~\bibnamefont {Su}}, \bibinfo {author} {\bibfnamefont {T.~J.}\ \bibnamefont {Marks}}, \bibinfo {author} {\bibfnamefont {M.~C.}\ \bibnamefont {Hersam}},\ and\ \bibinfo {author} {\bibfnamefont {E.~A.}\ \bibnamefont {Weiss}},\ }\bibfield  {title} {\bibinfo {title} {Molecular-orientation-dependent interfacial charge transfer in phthalocyanine/mos2 mixed-dimensional heterojunctions},\ }\href@noop {} {\bibfield  {journal} {\bibinfo  {journal} {The {J}ournal of {P}hysical {C}hemistry {C}}\ }\textbf {\bibinfo {volume} {123}},\ \bibinfo {pages} {13337} (\bibinfo {year} {2019})}\BibitemShut {NoStop}%
\bibitem [{\citenamefont {Padgaonkar}\ \emph {et~al.}(2020)\citenamefont {Padgaonkar}, \citenamefont {Olding}, \citenamefont {Lauhon}, \citenamefont {Hersam},\ and\ \citenamefont {Weiss}}]{padgaonkar2020emergent}%
  \BibitemOpen
  \bibfield  {author} {\bibinfo {author} {\bibfnamefont {S.}~\bibnamefont {Padgaonkar}}, \bibinfo {author} {\bibfnamefont {J.~N.}\ \bibnamefont {Olding}}, \bibinfo {author} {\bibfnamefont {L.~J.}\ \bibnamefont {Lauhon}}, \bibinfo {author} {\bibfnamefont {M.~C.}\ \bibnamefont {Hersam}},\ and\ \bibinfo {author} {\bibfnamefont {E.~A.}\ \bibnamefont {Weiss}},\ }\bibfield  {title} {\bibinfo {title} {Emergent optoelectronic properties of mixed-dimensional heterojunctions},\ }\href@noop {} {\bibfield  {journal} {\bibinfo  {journal} {Accounts of {C}hemical {R}esearch}\ }\textbf {\bibinfo {volume} {53}},\ \bibinfo {pages} {763} (\bibinfo {year} {2020})}\BibitemShut {NoStop}%
\bibitem [{\citenamefont {Li}\ \emph {et~al.}(2020)\citenamefont {Li}, \citenamefont {Zhong}, \citenamefont {Henning}, \citenamefont {Sangwan}, \citenamefont {Zhou}, \citenamefont {Liu}, \citenamefont {Rahn}, \citenamefont {Wells}, \citenamefont {Park}, \citenamefont {Luxa} \emph {et~al.}}]{li2020molecular}%
  \BibitemOpen
  \bibfield  {author} {\bibinfo {author} {\bibfnamefont {S.}~\bibnamefont {Li}}, \bibinfo {author} {\bibfnamefont {C.}~\bibnamefont {Zhong}}, \bibinfo {author} {\bibfnamefont {A.}~\bibnamefont {Henning}}, \bibinfo {author} {\bibfnamefont {V.~K.}\ \bibnamefont {Sangwan}}, \bibinfo {author} {\bibfnamefont {Q.}~\bibnamefont {Zhou}}, \bibinfo {author} {\bibfnamefont {X.}~\bibnamefont {Liu}}, \bibinfo {author} {\bibfnamefont {M.~S.}\ \bibnamefont {Rahn}}, \bibinfo {author} {\bibfnamefont {S.~A.}\ \bibnamefont {Wells}}, \bibinfo {author} {\bibfnamefont {H.~Y.}\ \bibnamefont {Park}}, \bibinfo {author} {\bibfnamefont {J.}~\bibnamefont {Luxa}}, \emph {et~al.},\ }\bibfield  {title} {\bibinfo {title} {Molecular-scale characterization of photoinduced charge separation in mixed-dimensional {InSe}--organic van der {W}aals heterostructures},\ }\href@noop {} {\bibfield  {journal} {\bibinfo  {journal} {{ACS} {N}ano}\ }\textbf {\bibinfo {volume} {14}},\ \bibinfo {pages} {3509} (\bibinfo {year} {2020})}\BibitemShut {NoStop}%
\bibitem [{\citenamefont {Amsterdam}\ \emph {et~al.}(2021{\natexlab{a}})\citenamefont {Amsterdam}, \citenamefont {Marks},\ and\ \citenamefont {Hersam}}]{amsterdam2021leveraging}%
  \BibitemOpen
  \bibfield  {author} {\bibinfo {author} {\bibfnamefont {S.~H.}\ \bibnamefont {Amsterdam}}, \bibinfo {author} {\bibfnamefont {T.~J.}\ \bibnamefont {Marks}},\ and\ \bibinfo {author} {\bibfnamefont {M.~C.}\ \bibnamefont {Hersam}},\ }\bibfield  {title} {\bibinfo {title} {Leveraging molecular properties to tailor mixed-dimensional heterostructures beyond energy level alignment},\ }\href@noop {} {\bibfield  {journal} {\bibinfo  {journal} {{J}ournal of {P}hysical {C}hemistry {L}etters}\ }\textbf {\bibinfo {volume} {12}},\ \bibinfo {pages} {4543} (\bibinfo {year} {2021}{\natexlab{a}})}\BibitemShut {NoStop}%
\bibitem [{\citenamefont {Utama}\ \emph {et~al.}(2023)\citenamefont {Utama}, \citenamefont {Dasgupta}, \citenamefont {Ananth}, \citenamefont {Weiss}, \citenamefont {Marks},\ and\ \citenamefont {Hersam}}]{utama2023mixed}%
  \BibitemOpen
  \bibfield  {author} {\bibinfo {author} {\bibfnamefont {M.~I.~B.}\ \bibnamefont {Utama}}, \bibinfo {author} {\bibfnamefont {A.}~\bibnamefont {Dasgupta}}, \bibinfo {author} {\bibfnamefont {R.}~\bibnamefont {Ananth}}, \bibinfo {author} {\bibfnamefont {E.~A.}\ \bibnamefont {Weiss}}, \bibinfo {author} {\bibfnamefont {T.~J.}\ \bibnamefont {Marks}},\ and\ \bibinfo {author} {\bibfnamefont {M.~C.}\ \bibnamefont {Hersam}},\ }\bibfield  {title} {\bibinfo {title} {Mixed-dimensional heterostructures for quantum photonic science and technology},\ }\href@noop {} {\bibfield  {journal} {\bibinfo  {journal} {{MRS} {B}ulletin}\ }\textbf {\bibinfo {volume} {48}},\ \bibinfo {pages} {905} (\bibinfo {year} {2023})}\BibitemShut {NoStop}%
\bibitem [{\citenamefont {Basova}\ \emph {et~al.}(2008)\citenamefont {Basova}, \citenamefont {Plyashkevich},\ and\ \citenamefont {Hassan}}]{basova2008spectral}%
  \BibitemOpen
  \bibfield  {author} {\bibinfo {author} {\bibfnamefont {T.}~\bibnamefont {Basova}}, \bibinfo {author} {\bibfnamefont {V.}~\bibnamefont {Plyashkevich}},\ and\ \bibinfo {author} {\bibfnamefont {A.}~\bibnamefont {Hassan}},\ }\bibfield  {title} {\bibinfo {title} {Spectral characterization of thin films of vanadyl hexadecafluorophthalocyanine {VOPcF}$_16$},\ }\href@noop {} {\bibfield  {journal} {\bibinfo  {journal} {Surface {S}cience}\ }\textbf {\bibinfo {volume} {602}},\ \bibinfo {pages} {2368} (\bibinfo {year} {2008})}\BibitemShut {NoStop}%
\bibitem [{\citenamefont {Nanai}\ \emph {et~al.}(1997)\citenamefont {Nanai}, \citenamefont {Yudasaka}, \citenamefont {Ohki},\ and\ \citenamefont {Yoshimura}}]{nanai1997polarized}%
  \BibitemOpen
  \bibfield  {author} {\bibinfo {author} {\bibfnamefont {N.}~\bibnamefont {Nanai}}, \bibinfo {author} {\bibfnamefont {M.}~\bibnamefont {Yudasaka}}, \bibinfo {author} {\bibfnamefont {Y.}~\bibnamefont {Ohki}},\ and\ \bibinfo {author} {\bibfnamefont {S.}~\bibnamefont {Yoshimura}},\ }\bibfield  {title} {\bibinfo {title} {Polarized optical absorption spectra of orientation aligned vanadyl phthalocyanine films},\ }\href@noop {} {\bibfield  {journal} {\bibinfo  {journal} {Thin {S}olid {F}ilms}\ }\textbf {\bibinfo {volume} {298}},\ \bibinfo {pages} {83} (\bibinfo {year} {1997})}\BibitemShut {NoStop}%
\bibitem [{\citenamefont {Wickramaratne}\ \emph {et~al.}(2018)\citenamefont {Wickramaratne}, \citenamefont {Weston},\ and\ \citenamefont {Van~de Walle}}]{wickramaratne2018monolayer}%
  \BibitemOpen
  \bibfield  {author} {\bibinfo {author} {\bibfnamefont {D.}~\bibnamefont {Wickramaratne}}, \bibinfo {author} {\bibfnamefont {L.}~\bibnamefont {Weston}},\ and\ \bibinfo {author} {\bibfnamefont {C.~G.}\ \bibnamefont {Van~de Walle}},\ }\bibfield  {title} {\bibinfo {title} {Monolayer to bulk properties of hexagonal boron nitride},\ }\href@noop {} {\bibfield  {journal} {\bibinfo  {journal} {The {J}ournal of {P}hysical {C}hemistry {C}}\ }\textbf {\bibinfo {volume} {122}},\ \bibinfo {pages} {25524} (\bibinfo {year} {2018})}\BibitemShut {NoStop}%
\bibitem [{\citenamefont {Schwinn}\ \emph {et~al.}(2022)\citenamefont {Schwinn}, \citenamefont {Rather}, \citenamefont {Lee}, \citenamefont {Bland}, \citenamefont {Song}, \citenamefont {Sangwan}, \citenamefont {Hersam},\ and\ \citenamefont {Chen}}]{schwinn2022charge}%
  \BibitemOpen
  \bibfield  {author} {\bibinfo {author} {\bibfnamefont {M.~C.}\ \bibnamefont {Schwinn}}, \bibinfo {author} {\bibfnamefont {S.~R.}\ \bibnamefont {Rather}}, \bibinfo {author} {\bibfnamefont {C.}~\bibnamefont {Lee}}, \bibinfo {author} {\bibfnamefont {M.~P.}\ \bibnamefont {Bland}}, \bibinfo {author} {\bibfnamefont {T.~W.}\ \bibnamefont {Song}}, \bibinfo {author} {\bibfnamefont {V.~K.}\ \bibnamefont {Sangwan}}, \bibinfo {author} {\bibfnamefont {M.~C.}\ \bibnamefont {Hersam}},\ and\ \bibinfo {author} {\bibfnamefont {L.~X.}\ \bibnamefont {Chen}},\ }\bibfield  {title} {\bibinfo {title} {Charge transfer dynamics and interlayer exciton formation in {MoS}$_2$/{VOPc} mixed dimensional heterojunction},\ }\href@noop {} {\bibfield  {journal} {\bibinfo  {journal} {The {J}ournal of {C}hemical {P}hysics}\ }\textbf {\bibinfo {volume} {157}} (\bibinfo {year} {2022})}\BibitemShut {NoStop}%
\bibitem [{\citenamefont {Kong}\ \emph {et~al.}(2022)\citenamefont {Kong}, \citenamefont {Obaidulla}, \citenamefont {Habib}, \citenamefont {Wang}, \citenamefont {Wang}, \citenamefont {Khan}, \citenamefont {Zhu}, \citenamefont {Xu},\ and\ \citenamefont {Yang}}]{kong2022interlayer}%
  \BibitemOpen
  \bibfield  {author} {\bibinfo {author} {\bibfnamefont {Y.}~\bibnamefont {Kong}}, \bibinfo {author} {\bibfnamefont {S.~M.}\ \bibnamefont {Obaidulla}}, \bibinfo {author} {\bibfnamefont {M.~R.}\ \bibnamefont {Habib}}, \bibinfo {author} {\bibfnamefont {Z.}~\bibnamefont {Wang}}, \bibinfo {author} {\bibfnamefont {R.}~\bibnamefont {Wang}}, \bibinfo {author} {\bibfnamefont {Y.}~\bibnamefont {Khan}}, \bibinfo {author} {\bibfnamefont {H.}~\bibnamefont {Zhu}}, \bibinfo {author} {\bibfnamefont {M.}~\bibnamefont {Xu}},\ and\ \bibinfo {author} {\bibfnamefont {D.}~\bibnamefont {Yang}},\ }\bibfield  {title} {\bibinfo {title} {Interlayer exciton emission in a {MoS} 2/{VOPc} inorganic/organic van der {W}aals heterostructure},\ }\href@noop {} {\bibfield  {journal} {\bibinfo  {journal} {Materials {H}orizons}\ }\textbf {\bibinfo {volume} {9}},\ \bibinfo {pages} {1253} (\bibinfo {year} {2022})}\BibitemShut {NoStop}%
\bibitem [{\citenamefont {Tomarchio}\ \emph {et~al.}(2021)\citenamefont {Tomarchio}, \citenamefont {Macis}, \citenamefont {Mosesso}, \citenamefont {Nguyen}, \citenamefont {Grilli}, \citenamefont {Guidi}, \citenamefont {Cava},\ and\ \citenamefont {Lupi}}]{tomarchio2021low}%
  \BibitemOpen
  \bibfield  {author} {\bibinfo {author} {\bibfnamefont {L.}~\bibnamefont {Tomarchio}}, \bibinfo {author} {\bibfnamefont {S.}~\bibnamefont {Macis}}, \bibinfo {author} {\bibfnamefont {L.}~\bibnamefont {Mosesso}}, \bibinfo {author} {\bibfnamefont {L.~T.}\ \bibnamefont {Nguyen}}, \bibinfo {author} {\bibfnamefont {A.}~\bibnamefont {Grilli}}, \bibinfo {author} {\bibfnamefont {M.~C.}\ \bibnamefont {Guidi}}, \bibinfo {author} {\bibfnamefont {R.~J.}\ \bibnamefont {Cava}},\ and\ \bibinfo {author} {\bibfnamefont {S.}~\bibnamefont {Lupi}},\ }\bibfield  {title} {\bibinfo {title} {Low energy electrodynamics of {CrI}$_3$ layered ferromagnet},\ }\href@noop {} {\bibfield  {journal} {\bibinfo  {journal} {Scientific {R}eports}\ }\textbf {\bibinfo {volume} {11}},\ \bibinfo {pages} {23405} (\bibinfo {year} {2021})}\BibitemShut {NoStop}%
\bibitem [{\citenamefont {Seyler}\ \emph {et~al.}(2018)\citenamefont {Seyler}, \citenamefont {Zhong}, \citenamefont {Klein}, \citenamefont {Gao}, \citenamefont {Zhang}, \citenamefont {Huang}, \citenamefont {Navarro-Moratalla}, \citenamefont {Yang}, \citenamefont {Cobden}, \citenamefont {McGuire} \emph {et~al.}}]{seyler2018ligand}%
  \BibitemOpen
  \bibfield  {author} {\bibinfo {author} {\bibfnamefont {K.~L.}\ \bibnamefont {Seyler}}, \bibinfo {author} {\bibfnamefont {D.}~\bibnamefont {Zhong}}, \bibinfo {author} {\bibfnamefont {D.~R.}\ \bibnamefont {Klein}}, \bibinfo {author} {\bibfnamefont {S.}~\bibnamefont {Gao}}, \bibinfo {author} {\bibfnamefont {X.}~\bibnamefont {Zhang}}, \bibinfo {author} {\bibfnamefont {B.}~\bibnamefont {Huang}}, \bibinfo {author} {\bibfnamefont {E.}~\bibnamefont {Navarro-Moratalla}}, \bibinfo {author} {\bibfnamefont {L.}~\bibnamefont {Yang}}, \bibinfo {author} {\bibfnamefont {D.~H.}\ \bibnamefont {Cobden}}, \bibinfo {author} {\bibfnamefont {M.~A.}\ \bibnamefont {McGuire}}, \emph {et~al.},\ }\bibfield  {title} {\bibinfo {title} {Ligand-field helical luminescence in a {2D} ferromagnetic insulator},\ }\href@noop {} {\bibfield  {journal} {\bibinfo  {journal} {Nature {P}hysics}\ }\textbf {\bibinfo {volume} {14}},\ \bibinfo {pages} {277} (\bibinfo {year} {2018})}\BibitemShut {NoStop}%
\bibitem [{\citenamefont {Koll}\ \emph {et~al.}(2024)\citenamefont {Koll}, \citenamefont {Urdaniz}, \citenamefont {Noh}, \citenamefont {Bae}, \citenamefont {Wolf},\ and\ \citenamefont {Gupta}}]{koll2024formation}%
  \BibitemOpen
  \bibfield  {author} {\bibinfo {author} {\bibfnamefont {W.}~\bibnamefont {Koll}}, \bibinfo {author} {\bibfnamefont {C.}~\bibnamefont {Urdaniz}}, \bibinfo {author} {\bibfnamefont {K.}~\bibnamefont {Noh}}, \bibinfo {author} {\bibfnamefont {Y.}~\bibnamefont {Bae}}, \bibinfo {author} {\bibfnamefont {C.}~\bibnamefont {Wolf}},\ and\ \bibinfo {author} {\bibfnamefont {J.}~\bibnamefont {Gupta}},\ }\bibfield  {title} {\bibinfo {title} {Formation of oriented bilayer motif: Vanadyl phthalocyanine on {Ag} (100)},\ }\href@noop {} {\bibfield  {journal} {\bibinfo  {journal} {The {J}ournal of {P}hysical {C}hemistry {C}}\ }\textbf {\bibinfo {volume} {128}},\ \bibinfo {pages} {12206} (\bibinfo {year} {2024})}\BibitemShut {NoStop}%
\bibitem [{\citenamefont {Kavand}\ \emph {et~al.}(2024)\citenamefont {Kavand}, \citenamefont {Phillips}, \citenamefont {Koll}, \citenamefont {Hamilton}, \citenamefont {Perez-Hoyos}, \citenamefont {Greer}, \citenamefont {Ara}, \citenamefont {Pharis}, \citenamefont {Maleki}, \citenamefont {Xu} \emph {et~al.}}]{kavand2024general}%
  \BibitemOpen
  \bibfield  {author} {\bibinfo {author} {\bibfnamefont {M.}~\bibnamefont {Kavand}}, \bibinfo {author} {\bibfnamefont {Z.}~\bibnamefont {Phillips}}, \bibinfo {author} {\bibfnamefont {W.~H.}\ \bibnamefont {Koll}}, \bibinfo {author} {\bibfnamefont {M.}~\bibnamefont {Hamilton}}, \bibinfo {author} {\bibfnamefont {E.}~\bibnamefont {Perez-Hoyos}}, \bibinfo {author} {\bibfnamefont {R.}~\bibnamefont {Greer}}, \bibinfo {author} {\bibfnamefont {F.}~\bibnamefont {Ara}}, \bibinfo {author} {\bibfnamefont {D.}~\bibnamefont {Pharis}}, \bibinfo {author} {\bibfnamefont {K.}~\bibnamefont {Maleki}}, \bibinfo {author} {\bibfnamefont {M.}~\bibnamefont {Xu}}, \emph {et~al.},\ }\bibfield  {title} {\bibinfo {title} {A general and modular approach to solid-state integration and readout of zero-dimensional quantum systems},\ }\href@noop {} {\bibfield  {journal} {\bibinfo  {journal} {{arXiv} preprint {arXiv}:2407.11189}\ } (\bibinfo {year} {2024})}\BibitemShut {NoStop}%
\bibitem [{\citenamefont {Amsterdam}\ \emph {et~al.}(2021{\natexlab{b}})\citenamefont {Amsterdam}, \citenamefont {Stanev}, \citenamefont {Wang}, \citenamefont {Zhou}, \citenamefont {Irgen-Gioro}, \citenamefont {Padgaonkar}, \citenamefont {Murthy}, \citenamefont {Sangwan}, \citenamefont {Dravid}, \citenamefont {Weiss} \emph {et~al.}}]{amsterdam2021mechanistic}%
  \BibitemOpen
  \bibfield  {author} {\bibinfo {author} {\bibfnamefont {S.~H.}\ \bibnamefont {Amsterdam}}, \bibinfo {author} {\bibfnamefont {T.~K.}\ \bibnamefont {Stanev}}, \bibinfo {author} {\bibfnamefont {L.}~\bibnamefont {Wang}}, \bibinfo {author} {\bibfnamefont {Q.}~\bibnamefont {Zhou}}, \bibinfo {author} {\bibfnamefont {S.}~\bibnamefont {Irgen-Gioro}}, \bibinfo {author} {\bibfnamefont {S.}~\bibnamefont {Padgaonkar}}, \bibinfo {author} {\bibfnamefont {A.~A.}\ \bibnamefont {Murthy}}, \bibinfo {author} {\bibfnamefont {V.~K.}\ \bibnamefont {Sangwan}}, \bibinfo {author} {\bibfnamefont {V.~P.}\ \bibnamefont {Dravid}}, \bibinfo {author} {\bibfnamefont {E.~A.}\ \bibnamefont {Weiss}}, \emph {et~al.},\ }\bibfield  {title} {\bibinfo {title} {Mechanistic investigation of molybdenum disulfide defect photoluminescence quenching by adsorbed metallophthalocyanines},\ }\href@noop {} {\bibfield  {journal} {\bibinfo  {journal} {Journal of the {A}merican {C}hemical {S}ociety}\ }\textbf {\bibinfo {volume} {143}},\ \bibinfo {pages} {17153}
  (\bibinfo {year} {2021}{\natexlab{b}})}\BibitemShut {NoStop}%
\bibitem [{\citenamefont {Rana}\ \emph {et~al.}(2022)\citenamefont {Rana}, \citenamefont {Johnson}, \citenamefont {Gurdumov}, \citenamefont {Mazur},\ and\ \citenamefont {Hipps}}]{rana2022scanning}%
  \BibitemOpen
  \bibfield  {author} {\bibinfo {author} {\bibfnamefont {S.}~\bibnamefont {Rana}}, \bibinfo {author} {\bibfnamefont {K.~N.}\ \bibnamefont {Johnson}}, \bibinfo {author} {\bibfnamefont {K.}~\bibnamefont {Gurdumov}}, \bibinfo {author} {\bibfnamefont {U.}~\bibnamefont {Mazur}},\ and\ \bibinfo {author} {\bibfnamefont {K.}~\bibnamefont {Hipps}},\ }\bibfield  {title} {\bibinfo {title} {Scanning tunneling microscopy reveals surface diffusion of single double-decker phthalocyanine molecules at the solution/solid interface},\ }\href@noop {} {\bibfield  {journal} {\bibinfo  {journal} {The Journal of {P}hysical {C}hemistry {C}}\ }\textbf {\bibinfo {volume} {126}},\ \bibinfo {pages} {4140} (\bibinfo {year} {2022})}\BibitemShut {NoStop}%
\bibitem [{\citenamefont {Furuyama}\ \emph {et~al.}(2020)\citenamefont {Furuyama}, \citenamefont {Uchiyama}, \citenamefont {Chikamatsu}, \citenamefont {Horikawa}, \citenamefont {Maeda}, \citenamefont {Segi}, \citenamefont {Takahashi},\ and\ \citenamefont {Taima}}]{furuyama2020temperature}%
  \BibitemOpen
  \bibfield  {author} {\bibinfo {author} {\bibfnamefont {T.}~\bibnamefont {Furuyama}}, \bibinfo {author} {\bibfnamefont {S.}~\bibnamefont {Uchiyama}}, \bibinfo {author} {\bibfnamefont {T.}~\bibnamefont {Chikamatsu}}, \bibinfo {author} {\bibfnamefont {T.}~\bibnamefont {Horikawa}}, \bibinfo {author} {\bibfnamefont {H.}~\bibnamefont {Maeda}}, \bibinfo {author} {\bibfnamefont {M.}~\bibnamefont {Segi}}, \bibinfo {author} {\bibfnamefont {H.}~\bibnamefont {Takahashi}},\ and\ \bibinfo {author} {\bibfnamefont {T.}~\bibnamefont {Taima}},\ }\bibfield  {title} {\bibinfo {title} {Temperature-dependent changes in the molecular orientation and visible color of phthalocyanine films},\ }\href@noop {} {\bibfield  {journal} {\bibinfo  {journal} {RSC {A}dvances}\ }\textbf {\bibinfo {volume} {10}},\ \bibinfo {pages} {31348} (\bibinfo {year} {2020})}\BibitemShut {NoStop}%
\bibitem [{\citenamefont {Kopaczek}\ \emph {et~al.}(2022)\citenamefont {Kopaczek}, \citenamefont {Zelewski}, \citenamefont {Yumigeta}, \citenamefont {Sailus}, \citenamefont {Tongay},\ and\ \citenamefont {Kudrawiec}}]{kopaczek2022temperature}%
  \BibitemOpen
  \bibfield  {author} {\bibinfo {author} {\bibfnamefont {J.}~\bibnamefont {Kopaczek}}, \bibinfo {author} {\bibfnamefont {S.}~\bibnamefont {Zelewski}}, \bibinfo {author} {\bibfnamefont {K.}~\bibnamefont {Yumigeta}}, \bibinfo {author} {\bibfnamefont {R.}~\bibnamefont {Sailus}}, \bibinfo {author} {\bibfnamefont {S.}~\bibnamefont {Tongay}},\ and\ \bibinfo {author} {\bibfnamefont {R.}~\bibnamefont {Kudrawiec}},\ }\bibfield  {title} {\bibinfo {title} {Temperature dependence of the indirect gap and the direct optical transitions at the high-symmetry point of the brillouin zone and band nesting in {MoS}$_2$, {MoSe}$_2$, {MoTe}$_2$, {WS}$_2$, and {WSe}$_2$ crystals},\ }\href@noop {} {\bibfield  {journal} {\bibinfo  {journal} {The {J}ournal of {P}hysical {C}hemistry {C}}\ }\textbf {\bibinfo {volume} {126}},\ \bibinfo {pages} {5665} (\bibinfo {year} {2022})}\BibitemShut {NoStop}%
\bibitem [{\citenamefont {Pan}\ \emph {et~al.}(1998)\citenamefont {Pan}, \citenamefont {Wu}, \citenamefont {Chen}, \citenamefont {Zhao}, \citenamefont {Shen}, \citenamefont {Li}, \citenamefont {Shen},\ and\ \citenamefont {Huang}}]{pan1998structure}%
  \BibitemOpen
  \bibfield  {author} {\bibinfo {author} {\bibfnamefont {Y.}~\bibnamefont {Pan}}, \bibinfo {author} {\bibfnamefont {Y.}~\bibnamefont {Wu}}, \bibinfo {author} {\bibfnamefont {L.}~\bibnamefont {Chen}}, \bibinfo {author} {\bibfnamefont {Y.}~\bibnamefont {Zhao}}, \bibinfo {author} {\bibfnamefont {Y.}~\bibnamefont {Shen}}, \bibinfo {author} {\bibfnamefont {F.}~\bibnamefont {Li}}, \bibinfo {author} {\bibfnamefont {S.}~\bibnamefont {Shen}},\ and\ \bibinfo {author} {\bibfnamefont {D.}~\bibnamefont {Huang}},\ }\bibfield  {title} {\bibinfo {title} {Structure and spectroscopic characterization of polycrystalline vanadyl phthalocyanine ({VOPc}) films fabricated by vacuum deposition.},\ }\href@noop {} {\bibfield  {journal} {\bibinfo  {journal} {Applied {P}hysics {A}: {M}aterials {S}cience \& {P}rocessing}\ }\textbf {\bibinfo {volume} {66}} (\bibinfo {year} {1998})}\BibitemShut {NoStop}%
\bibitem [{\citenamefont {Singh}\ and\ \citenamefont {Ravindra}(2010)}]{singh2010optical}%
  \BibitemOpen
  \bibfield  {author} {\bibinfo {author} {\bibfnamefont {P.}~\bibnamefont {Singh}}\ and\ \bibinfo {author} {\bibfnamefont {N.}~\bibnamefont {Ravindra}},\ }\bibfield  {title} {\bibinfo {title} {Optical properties of metal phthalocyanines},\ }\href@noop {} {\bibfield  {journal} {\bibinfo  {journal} {Journal of {M}aterials {S}cience}\ }\textbf {\bibinfo {volume} {45}},\ \bibinfo {pages} {4013} (\bibinfo {year} {2010})}\BibitemShut {NoStop}%
\bibitem [{\citenamefont {Baba}\ \emph {et~al.}(2016)\citenamefont {Baba}, \citenamefont {Suzuki},\ and\ \citenamefont {Oku}}]{baba2016electronic}%
  \BibitemOpen
  \bibfield  {author} {\bibinfo {author} {\bibfnamefont {S.}~\bibnamefont {Baba}}, \bibinfo {author} {\bibfnamefont {A.}~\bibnamefont {Suzuki}},\ and\ \bibinfo {author} {\bibfnamefont {T.}~\bibnamefont {Oku}},\ }\bibfield  {title} {\bibinfo {title} {Electronic structures and magnetic/optical properties of metal phthalocyanine complexes},\ }in\ \href@noop {} {\emph {\bibinfo {booktitle} {{AIP} {C}onference {P}roceedings}}},\ Vol.\ \bibinfo {volume} {1709}\ (\bibinfo {organization} {AIP Publishing},\ \bibinfo {year} {2016})\BibitemShut {NoStop}%
\bibitem [{\citenamefont {Meletov}\ \emph {et~al.}(2020)\citenamefont {Meletov}, \citenamefont {Kuzmin},\ and\ \citenamefont {Shibaeva}}]{meletov2020phonon}%
  \BibitemOpen
  \bibfield  {author} {\bibinfo {author} {\bibfnamefont {K.}~\bibnamefont {Meletov}}, \bibinfo {author} {\bibfnamefont {A.}~\bibnamefont {Kuzmin}},\ and\ \bibinfo {author} {\bibfnamefont {R.}~\bibnamefont {Shibaeva}},\ }\bibfield  {title} {\bibinfo {title} {Phonon spectrum and structural transformations at high pressures in vanadyl {IV} phthalocyanine crystals},\ }\href@noop {} {\bibfield  {journal} {\bibinfo  {journal} {Journal of {E}xperimental and {T}heoretical {P}hysics}\ }\textbf {\bibinfo {volume} {130}},\ \bibinfo {pages} {69} (\bibinfo {year} {2020})}\BibitemShut {NoStop}%
\bibitem [{\citenamefont {Xiong}\ \emph {et~al.}(2024)\citenamefont {Xiong}, \citenamefont {Wang}, \citenamefont {Yao}, \citenamefont {Xu},\ and\ \citenamefont {Xu}}]{xiong2024exciton}%
  \BibitemOpen
  \bibfield  {author} {\bibinfo {author} {\bibfnamefont {S.}~\bibnamefont {Xiong}}, \bibinfo {author} {\bibfnamefont {Y.}~\bibnamefont {Wang}}, \bibinfo {author} {\bibfnamefont {J.}~\bibnamefont {Yao}}, \bibinfo {author} {\bibfnamefont {J.}~\bibnamefont {Xu}},\ and\ \bibinfo {author} {\bibfnamefont {M.}~\bibnamefont {Xu}},\ }\bibfield  {title} {\bibinfo {title} {Exciton dynamics of {TiOPc}/{WSe}$_2$ heterostructure},\ }\href@noop {} {\bibfield  {journal} {\bibinfo  {journal} {ACS {N}ano}\ }\textbf {\bibinfo {volume} {18}},\ \bibinfo {pages} {10249} (\bibinfo {year} {2024})}\BibitemShut {NoStop}%
\bibitem [{\citenamefont {Wan}\ \emph {et~al.}(2021)\citenamefont {Wan}, \citenamefont {Wang}, \citenamefont {Liu},\ and\ \citenamefont {Liang}}]{wan2021effectively}%
  \BibitemOpen
  \bibfield  {author} {\bibinfo {author} {\bibfnamefont {Z.}~\bibnamefont {Wan}}, \bibinfo {author} {\bibfnamefont {Q.-D.}\ \bibnamefont {Wang}}, \bibinfo {author} {\bibfnamefont {D.}~\bibnamefont {Liu}},\ and\ \bibinfo {author} {\bibfnamefont {J.}~\bibnamefont {Liang}},\ }\bibfield  {title} {\bibinfo {title} {Effectively improving the accuracy of {PBE} functional in calculating the solid band gap via machine learning},\ }\href@noop {} {\bibfield  {journal} {\bibinfo  {journal} {Computational {M}aterials {S}cience}\ }\textbf {\bibinfo {volume} {198}},\ \bibinfo {pages} {110699} (\bibinfo {year} {2021})}\BibitemShut {NoStop}%
\bibitem [{\citenamefont {Zhang}\ \emph {et~al.}(2007)\citenamefont {Zhang}, \citenamefont {Learmonth}, \citenamefont {Wang}, \citenamefont {Matsuura}, \citenamefont {Downes}, \citenamefont {Plucinski}, \citenamefont {Bernardis}, \citenamefont {O'Donnell},\ and\ \citenamefont {Smith}}]{zhang2007electronic}%
  \BibitemOpen
  \bibfield  {author} {\bibinfo {author} {\bibfnamefont {Y.}~\bibnamefont {Zhang}}, \bibinfo {author} {\bibfnamefont {T.}~\bibnamefont {Learmonth}}, \bibinfo {author} {\bibfnamefont {S.}~\bibnamefont {Wang}}, \bibinfo {author} {\bibfnamefont {A.~Y.}\ \bibnamefont {Matsuura}}, \bibinfo {author} {\bibfnamefont {J.}~\bibnamefont {Downes}}, \bibinfo {author} {\bibfnamefont {L.}~\bibnamefont {Plucinski}}, \bibinfo {author} {\bibfnamefont {S.}~\bibnamefont {Bernardis}}, \bibinfo {author} {\bibfnamefont {C.}~\bibnamefont {O'Donnell}},\ and\ \bibinfo {author} {\bibfnamefont {K.~E.}\ \bibnamefont {Smith}},\ }\bibfield  {title} {\bibinfo {title} {Electronic structure of the organic semiconductor vanadyl phthalocyanine ({VO}-{P}c)},\ }\href@noop {} {\bibfield  {journal} {\bibinfo  {journal} {Journal of {M}aterials {C}hemistry}\ } (\bibinfo {year} {2007})}\BibitemShut {NoStop}%
\bibitem [{\citenamefont {Debnath}\ \emph {et~al.}(2022)\citenamefont {Debnath}, \citenamefont {Haupa}, \citenamefont {Lebedkin}, \citenamefont {Strelnikov},\ and\ \citenamefont {Kappes}}]{debnath2022triggering}%
  \BibitemOpen
  \bibfield  {author} {\bibinfo {author} {\bibfnamefont {S.}~\bibnamefont {Debnath}}, \bibinfo {author} {\bibfnamefont {K.~A.}\ \bibnamefont {Haupa}}, \bibinfo {author} {\bibfnamefont {S.}~\bibnamefont {Lebedkin}}, \bibinfo {author} {\bibfnamefont {D.}~\bibnamefont {Strelnikov}},\ and\ \bibinfo {author} {\bibfnamefont {M.~M.}\ \bibnamefont {Kappes}},\ }\bibfield  {title} {\bibinfo {title} {Triggering near-infrared luminescence of vanadyl phthalocyanine by charging},\ }\href@noop {} {\bibfield  {journal} {\bibinfo  {journal} {Angewandte {C}hemie}\ }\textbf {\bibinfo {volume} {134}},\ \bibinfo {pages} {e202201577} (\bibinfo {year} {2022})}\BibitemShut {NoStop}%
\bibitem [{\citenamefont {Adler}\ \emph {et~al.}(2015)\citenamefont {Adler}, \citenamefont {Paszkiewicz}, \citenamefont {Uihlein}, \citenamefont {Polek}, \citenamefont {Ovsyannikov}, \citenamefont {Basova}, \citenamefont {Chassé},\ and\ \citenamefont {Peisert}}]{adler2015interface}%
  \BibitemOpen
  \bibfield  {author} {\bibinfo {author} {\bibfnamefont {H.}~\bibnamefont {Adler}}, \bibinfo {author} {\bibfnamefont {M.}~\bibnamefont {Paszkiewicz}}, \bibinfo {author} {\bibfnamefont {J.}~\bibnamefont {Uihlein}}, \bibinfo {author} {\bibfnamefont {M.}~\bibnamefont {Polek}}, \bibinfo {author} {\bibfnamefont {R.}~\bibnamefont {Ovsyannikov}}, \bibinfo {author} {\bibfnamefont {T.~V.}\ \bibnamefont {Basova}}, \bibinfo {author} {\bibfnamefont {T.}~\bibnamefont {Chassé}},\ and\ \bibinfo {author} {\bibfnamefont {H.}~\bibnamefont {Peisert}},\ }\bibfield  {title} {\bibinfo {title} {Interface properties of {VOPc} on {N}i (111) and graphene/{N}i (111): {O}rientation-dependent charge transfer},\ }\href@noop {} {\bibfield  {journal} {\bibinfo  {journal} {Journal of {P}hysical {C}hemistry {C}}\ }\textbf {\bibinfo {volume} {119}},\ \bibinfo {pages} {8755} (\bibinfo {year} {2015})}\BibitemShut {NoStop}%
\bibitem [{\citenamefont {Theisen}\ \emph {et~al.}(2015)\citenamefont {Theisen}, \citenamefont {Huang}, \citenamefont {Fleetham}, \citenamefont {Adams},\ and\ \citenamefont {Li}}]{theisen2015ground}%
  \BibitemOpen
  \bibfield  {author} {\bibinfo {author} {\bibfnamefont {R.~F.}\ \bibnamefont {Theisen}}, \bibinfo {author} {\bibfnamefont {L.}~\bibnamefont {Huang}}, \bibinfo {author} {\bibfnamefont {T.}~\bibnamefont {Fleetham}}, \bibinfo {author} {\bibfnamefont {J.~B.}\ \bibnamefont {Adams}},\ and\ \bibinfo {author} {\bibfnamefont {J.}~\bibnamefont {Li}},\ }\bibfield  {title} {\bibinfo {title} {Ground and excited states of zinc phthalocyanine, zinc tetrabenzoporphyrin, and azaporphyrin analogs using {DFT} and {TDDFT} with {F}ranck-{C}ondon analysis},\ }\href@noop {} {\bibfield  {journal} {\bibinfo  {journal} {The {J}ournal of {C}hemical {P}hysics}\ }\textbf {\bibinfo {volume} {142}} (\bibinfo {year} {2015})}\BibitemShut {NoStop}%
\bibitem [{\citenamefont {Wang}\ and\ \citenamefont {Ishii}(2022)}]{wang2022photochemical}%
  \BibitemOpen
  \bibfield  {author} {\bibinfo {author} {\bibfnamefont {M.}~\bibnamefont {Wang}}\ and\ \bibinfo {author} {\bibfnamefont {K.}~\bibnamefont {Ishii}},\ }\bibfield  {title} {\bibinfo {title} {Photochemical properties of phthalocyanines with transition metal ions},\ }\href@noop {} {\bibfield  {journal} {\bibinfo  {journal} {Coordination {C}hemistry {R}eviews}\ }\textbf {\bibinfo {volume} {468}},\ \bibinfo {pages} {214626} (\bibinfo {year} {2022})}\BibitemShut {NoStop}%
\bibitem [{\citenamefont {Cao}\ \emph {et~al.}(2022)\citenamefont {Cao}, \citenamefont {Gong}, \citenamefont {Wang}, \citenamefont {Tang}, \citenamefont {Wang},\ and\ \citenamefont {Zheng}}]{cao2022steady}%
  \BibitemOpen
  \bibfield  {author} {\bibinfo {author} {\bibfnamefont {H.}~\bibnamefont {Cao}}, \bibinfo {author} {\bibfnamefont {M.}~\bibnamefont {Gong}}, \bibinfo {author} {\bibfnamefont {M.}~\bibnamefont {Wang}}, \bibinfo {author} {\bibfnamefont {Q.}~\bibnamefont {Tang}}, \bibinfo {author} {\bibfnamefont {L.}~\bibnamefont {Wang}},\ and\ \bibinfo {author} {\bibfnamefont {X.}~\bibnamefont {Zheng}},\ }\bibfield  {title} {\bibinfo {title} {Steady/transient state spectral researches on the solvent-triggered and photo-induced novel properties of metal-coordinated phthalocyanines},\ }\href@noop {} {\bibfield  {journal} {\bibinfo  {journal} {RSC {A}dvances}\ }\textbf {\bibinfo {volume} {12}},\ \bibinfo {pages} {5964} (\bibinfo {year} {2022})}\BibitemShut {NoStop}%
\bibitem [{\citenamefont {Van~Mingroot}\ \emph {et~al.}(1996)\citenamefont {Van~Mingroot}, \citenamefont {De~Backer}, \citenamefont {van Stam}, \citenamefont {Van~der Auweraer},\ and\ \citenamefont {De~Schryver}}]{van1996emission}%
  \BibitemOpen
  \bibfield  {author} {\bibinfo {author} {\bibfnamefont {H.}~\bibnamefont {Van~Mingroot}}, \bibinfo {author} {\bibfnamefont {S.}~\bibnamefont {De~Backer}}, \bibinfo {author} {\bibfnamefont {J.}~\bibnamefont {van Stam}}, \bibinfo {author} {\bibfnamefont {M.}~\bibnamefont {Van~der Auweraer}},\ and\ \bibinfo {author} {\bibfnamefont {F.}~\bibnamefont {De~Schryver}},\ }\bibfield  {title} {\bibinfo {title} {The emission at 669 nm of metal free phthalocyanine in toluene and 1-bromonaphthalene solutions},\ }\href@noop {} {\bibfield  {journal} {\bibinfo  {journal} {Chemical {P}hysics {L}etters}\ }\textbf {\bibinfo {volume} {253}},\ \bibinfo {pages} {397} (\bibinfo {year} {1996})}\BibitemShut {NoStop}%
\bibitem [{\citenamefont {Edwards}\ and\ \citenamefont {Gouterman}(1970)}]{edwards1970porphyrins}%
  \BibitemOpen
  \bibfield  {author} {\bibinfo {author} {\bibfnamefont {L.}~\bibnamefont {Edwards}}\ and\ \bibinfo {author} {\bibfnamefont {M.}~\bibnamefont {Gouterman}},\ }\bibfield  {title} {\bibinfo {title} {Porphyrins: {XV}. {V}apor absorption spectra and stability: {P}hthalocyanines},\ }\href@noop {} {\bibfield  {journal} {\bibinfo  {journal} {Journal of {M}olecular {S}pectroscopy}\ }\textbf {\bibinfo {volume} {33}},\ \bibinfo {pages} {292} (\bibinfo {year} {1970})}\BibitemShut {NoStop}%
\bibitem [{\citenamefont {Veiga}\ \emph {et~al.}(2016)\citenamefont {Veiga}, \citenamefont {Miwa},\ and\ \citenamefont {McLean}}]{veiga2016adsorption}%
  \BibitemOpen
  \bibfield  {author} {\bibinfo {author} {\bibfnamefont {R.}~\bibnamefont {Veiga}}, \bibinfo {author} {\bibfnamefont {R.}~\bibnamefont {Miwa}},\ and\ \bibinfo {author} {\bibfnamefont {A.}~\bibnamefont {McLean}},\ }\bibfield  {title} {\bibinfo {title} {Adsorption of metal-phthalocyanine molecules onto the {Si} (111) surface passivated by $\delta$ doping: Ab initio calculations},\ }\href@noop {} {\bibfield  {journal} {\bibinfo  {journal} {Physical {R}eview {B}}\ }\textbf {\bibinfo {volume} {93}},\ \bibinfo {pages} {115301} (\bibinfo {year} {2016})}\BibitemShut {NoStop}%
\bibitem [{\citenamefont {Jarvinen}\ \emph {et~al.}(2013)\citenamefont {Jarvinen}, \citenamefont {Hamalainen}, \citenamefont {Banerjee}, \citenamefont {Hakkinen}, \citenamefont {Ijas}, \citenamefont {Harju},\ and\ \citenamefont {Liljeroth}}]{jarvinen2013molecular}%
  \BibitemOpen
  \bibfield  {author} {\bibinfo {author} {\bibfnamefont {P.}~\bibnamefont {Jarvinen}}, \bibinfo {author} {\bibfnamefont {S.~K.}\ \bibnamefont {Hamalainen}}, \bibinfo {author} {\bibfnamefont {K.}~\bibnamefont {Banerjee}}, \bibinfo {author} {\bibfnamefont {P.}~\bibnamefont {Hakkinen}}, \bibinfo {author} {\bibfnamefont {M.}~\bibnamefont {Ijas}}, \bibinfo {author} {\bibfnamefont {A.}~\bibnamefont {Harju}},\ and\ \bibinfo {author} {\bibfnamefont {P.}~\bibnamefont {Liljeroth}},\ }\bibfield  {title} {\bibinfo {title} {Molecular self-assembly on graphene on {SiO}$_2$ and h-{BN} substrates},\ }\href@noop {} {\bibfield  {journal} {\bibinfo  {journal} {Nano {L}etters}\ }\textbf {\bibinfo {volume} {13}},\ \bibinfo {pages} {3199} (\bibinfo {year} {2013})}\BibitemShut {NoStop}%
\bibitem [{\citenamefont {Jarvinen}\ \emph {et~al.}(2014)\citenamefont {Jarvinen}, \citenamefont {Hamalainen}, \citenamefont {Ijas}, \citenamefont {Harju},\ and\ \citenamefont {Liljeroth}}]{jarvinen2014self}%
  \BibitemOpen
  \bibfield  {author} {\bibinfo {author} {\bibfnamefont {P.}~\bibnamefont {Jarvinen}}, \bibinfo {author} {\bibfnamefont {S.~K.}\ \bibnamefont {Hamalainen}}, \bibinfo {author} {\bibfnamefont {M.}~\bibnamefont {Ijas}}, \bibinfo {author} {\bibfnamefont {A.}~\bibnamefont {Harju}},\ and\ \bibinfo {author} {\bibfnamefont {P.}~\bibnamefont {Liljeroth}},\ }\bibfield  {title} {\bibinfo {title} {Self-assembly and orbital imaging of metal phthalocyanines on a graphene model surface},\ }\href@noop {} {\bibfield  {journal} {\bibinfo  {journal} {The {J}ournal of {P}hysical {C}hemistry {C}}\ }\textbf {\bibinfo {volume} {118}},\ \bibinfo {pages} {13320} (\bibinfo {year} {2014})}\BibitemShut {NoStop}%
\bibitem [{\citenamefont {Tomoda}\ \emph {et~al.}(1983)\citenamefont {Tomoda}, \citenamefont {Saito},\ and\ \citenamefont {Shiraishi}}]{tomoda_synthesis_1983}%
  \BibitemOpen
  \bibfield  {author} {\bibinfo {author} {\bibfnamefont {H.}~\bibnamefont {Tomoda}}, \bibinfo {author} {\bibfnamefont {S.}~\bibnamefont {Saito}},\ and\ \bibinfo {author} {\bibfnamefont {S.}~\bibnamefont {Shiraishi}},\ }\bibfield  {title} {\bibinfo {title} {Synthesis of metallophthalocyanines from phthalonitrile with strong organic bases},\ }\href {https://doi.org/10.1246/cl.1983.313} {\bibfield  {journal} {\bibinfo  {journal} {Chemistry {L}etters}\ }\textbf {\bibinfo {volume} {12}},\ \bibinfo {pages} {313} (\bibinfo {year} {1983})}\BibitemShut {NoStop}%
\bibitem [{\citenamefont {Kresse}\ and\ \citenamefont {Joubert}(1999)}]{kresse1999ultrasoft}%
  \BibitemOpen
  \bibfield  {author} {\bibinfo {author} {\bibfnamefont {G.}~\bibnamefont {Kresse}}\ and\ \bibinfo {author} {\bibfnamefont {D.}~\bibnamefont {Joubert}},\ }\bibfield  {title} {\bibinfo {title} {From ultrasoft pseudopotentials to the projector augmented-wave method},\ }\href@noop {} {\bibfield  {journal} {\bibinfo  {journal} {Physical {R}eview {B}}\ }\textbf {\bibinfo {volume} {59}},\ \bibinfo {pages} {1758} (\bibinfo {year} {1999})}\BibitemShut {NoStop}%
\bibitem [{\citenamefont {Paier}\ \emph {et~al.}(2005)\citenamefont {Paier}, \citenamefont {Hirschl}, \citenamefont {Marsman},\ and\ \citenamefont {Kresse}}]{paier2005perdew}%
  \BibitemOpen
  \bibfield  {author} {\bibinfo {author} {\bibfnamefont {J.}~\bibnamefont {Paier}}, \bibinfo {author} {\bibfnamefont {R.}~\bibnamefont {Hirschl}}, \bibinfo {author} {\bibfnamefont {M.}~\bibnamefont {Marsman}},\ and\ \bibinfo {author} {\bibfnamefont {G.}~\bibnamefont {Kresse}},\ }\bibfield  {title} {\bibinfo {title} {The {P}erdew--{B}urke--{E}rnzerhof exchange-correlation functional applied to the g2-1 test set using a plane-wave basis set},\ }\href@noop {} {\bibfield  {journal} {\bibinfo  {journal} {The Journal of {C}hemical {P}hysics}\ }\textbf {\bibinfo {volume} {122}} (\bibinfo {year} {2005})}\BibitemShut {NoStop}%
\bibitem [{\citenamefont {Grimme}\ \emph {et~al.}(2010)\citenamefont {Grimme}, \citenamefont {Antony}, \citenamefont {Ehrlich},\ and\ \citenamefont {Krieg}}]{grimme2010consistent}%
  \BibitemOpen
  \bibfield  {author} {\bibinfo {author} {\bibfnamefont {S.}~\bibnamefont {Grimme}}, \bibinfo {author} {\bibfnamefont {J.}~\bibnamefont {Antony}}, \bibinfo {author} {\bibfnamefont {S.}~\bibnamefont {Ehrlich}},\ and\ \bibinfo {author} {\bibfnamefont {H.}~\bibnamefont {Krieg}},\ }\bibfield  {title} {\bibinfo {title} {A consistent and accurate ab initio parametrization of density functional dispersion correction ({DFT-D}) for the 94 elements {H}-{P}u},\ }\href@noop {} {\bibfield  {journal} {\bibinfo  {journal} {The {J}ournal of {C}hemical {P}hysics}\ }\textbf {\bibinfo {volume} {132}} (\bibinfo {year} {2010})}\BibitemShut {NoStop}%
\bibitem [{\citenamefont {Heyd}\ \emph {et~al.}(2003)\citenamefont {Heyd}, \citenamefont {Scuseria},\ and\ \citenamefont {Ernzerhof}}]{heyd2003hybrid}%
  \BibitemOpen
  \bibfield  {author} {\bibinfo {author} {\bibfnamefont {J.}~\bibnamefont {Heyd}}, \bibinfo {author} {\bibfnamefont {G.~E.}\ \bibnamefont {Scuseria}},\ and\ \bibinfo {author} {\bibfnamefont {M.}~\bibnamefont {Ernzerhof}},\ }\bibfield  {title} {\bibinfo {title} {Hybrid functionals based on a screened {C}oulomb potential},\ }\href@noop {} {\bibfield  {journal} {\bibinfo  {journal} {The Journal of {C}hemical {P}hysics}\ }\textbf {\bibinfo {volume} {118}},\ \bibinfo {pages} {8207} (\bibinfo {year} {2003})}\BibitemShut {NoStop}%
\bibitem [{\citenamefont {Wang}\ \emph {et~al.}(2021)\citenamefont {Wang}, \citenamefont {Xu}, \citenamefont {Liu}, \citenamefont {Tang},\ and\ \citenamefont {Geng}}]{wang2021vaspkit}%
  \BibitemOpen
  \bibfield  {author} {\bibinfo {author} {\bibfnamefont {V.}~\bibnamefont {Wang}}, \bibinfo {author} {\bibfnamefont {N.}~\bibnamefont {Xu}}, \bibinfo {author} {\bibfnamefont {J.-C.}\ \bibnamefont {Liu}}, \bibinfo {author} {\bibfnamefont {G.}~\bibnamefont {Tang}},\ and\ \bibinfo {author} {\bibfnamefont {W.-T.}\ \bibnamefont {Geng}},\ }\bibfield  {title} {\bibinfo {title} {Vaspkit: A user-friendly interface facilitating high-throughput computing and analysis using {VASP} code},\ }\href@noop {} {\bibfield  {journal} {\bibinfo  {journal} {Computer {P}hysics {C}ommunications}\ }\textbf {\bibinfo {volume} {267}},\ \bibinfo {pages} {108033} (\bibinfo {year} {2021})}\BibitemShut {NoStop}%
\bibitem [{\citenamefont {Di~Pietro}\ and\ \citenamefont {Kerridge}(2016)}]{di2016u}%
  \BibitemOpen
  \bibfield  {author} {\bibinfo {author} {\bibfnamefont {P.}~\bibnamefont {Di~Pietro}}\ and\ \bibinfo {author} {\bibfnamefont {A.}~\bibnamefont {Kerridge}},\ }\bibfield  {title} {\bibinfo {title} {U--oyl stretching vibrations as a quantitative measure of the equatorial bond covalency in uranyl complexes: a quantum-chemical investigation},\ }\href@noop {} {\bibfield  {journal} {\bibinfo  {journal} {Inorganic {C}hemistry}\ }\textbf {\bibinfo {volume} {55}},\ \bibinfo {pages} {573} (\bibinfo {year} {2016})}\BibitemShut {NoStop}%
\end{thebibliography}%

\end{document}